\documentclass[%
reprint,
 amsmath,amssymb,
 aps,
pra,
]{revtex4-2}

\usepackage[english]{babel}

\usepackage[utf8]{inputenc}
\usepackage{csquotes}
\usepackage{graphicx}
\usepackage{dcolumn}
\usepackage{bm}
\usepackage{siunitx}
\usepackage[version=4]{mhchem}
\usepackage{booktabs} 
\usepackage{threeparttable}

\newcommand{\Qi}{Q_\mathrm{i}}

\newcommand{\Qm}{Q_\mathrm{m}}
\newcommand{\Qmu}{Q_\mathrm{\mu}}
\newcommand{\gem}{g_\mathrm{em}}
\newcommand{\gammac}{\gamma_\mathrm{c}}
\newcommand{\gammai}{\gamma_\mathrm{i}}
\newcommand{\kappamu}{\kappa_\mathrm{\mu}}
\newcommand{\kappac}{\kappa_\mathrm{c}}
\newcommand{\kappai}{\kappa_\mathrm{i}}
\newcommand{\omegam}{\omega_\mathrm{m}}
\newcommand{\omegamu}{\omega_\mathrm{\mu}}
\newcommand{\CIDT}{C_\mathrm{IDT}}
\newcommand{\Cm}{C_\mathrm{m}}
\newcommand{\Cmu}{C_\mathrm{\mu}}

\newcommand{\Zmu}{Z_\mathrm{\mu}}

\begin{document}
%

\title{Release-free phononic crystal with strong microwave coupling}

\author{Joey Frey}
\email{joey.frey@chalmers.se}
\author{Paul Burger}
\author{Trond Hjerpekj{\o}n Haug}
\author{Johan Kolvik}
\author{Rapha\"{e}l Van Laer}
 \email{raphael.van.laer@chalmers.se}
\affiliation{Department of Microtechnology and Nanoscience (MC2), Chalmers University of Technology.}

\date{\today}
\begin{abstract}
Phonons hold promise for storing and transferring quantum information, including in mechanically-mediated quantum interconnects between superconducting qubits and light. Phononic crystal cavities confine gigahertz sound to micron-scale volumes well matched to near-infrared light. So far, these devices have typically been suspended to suppress phononic radiation loss into the substrate, but suspension limits thermal anchoring leading to excess noise. Release-free phononic crystals have emerged as a way to address this challenge -- but had yet to be shown compatible with strong electromechanical interactions. Here, we demonstrate a release-free phononic crystal cavity strongly coupled to a high-impedance microwave resonator, with an electromechanical coupling rate $\gem/(2\pi) \approx 30\,\text{MHz}$ that exceeds both the mechanical and microwave loss rates, leading to a cooperativity up to $\mathcal{C} \approx 180$ on resonance. In addition, our lithium niobate phononic crystals reach quality factors above $10^4$ at millikelvin temperature on both silicon and sapphire substrates. Our results establish release-free phononic crystals as compact, scalable interfaces between microwaves and gigahertz sound for emerging sensing, communication, and computing systems.
\end{abstract}
\maketitle
\section{Introduction}
\label{sec:introduction}
\label{int:motivation}
Phonons are promising carriers of quantum information. Confined to micron-scale volumes at gigahertz frequencies, they couple efficiently to optical photons \cite{chan_laser_2011}, microwave photons \cite{oconnell_quantum_2010, teufel_circuit_2011}, spins \cite{meesala_enhanced_2016}, and magnons \cite{hwang_strongly_2024}, making them a natural intermediary in hybrid quantum systems. In particular, gigahertz phonons offer a route to bridging superconducting microwave qubits and near-infrared light, enabling mechanically-mediated quantum links between remote superconducting processors \cite{mirhosseini_superconducting_2020}.

\label{int:suspension}
The wavelength-scale phononic crystal cavity is a central tool in this program \cite{eichenfield_optomechanical_2009,safavi-naeini_controlling_2019}. Patterning a thin film with a periodic structure supporting a phononic band gap lets mechanical motion be co-confined with optical and microwave fields with high overlap and a small footprint \cite{chan_optimized_2012, safavi-naeini_design_2010, eichenfield_optomechanical_2009}. Built on suspended thin films, these devices have enabled several key advances, including mechanical quality factors as high as $10^{10}$ \cite{maccabe_nano-acoustic_2020}, mechanical ground-state cooling \cite{chan_laser_2011}, and quantum experiments combining microwave and optical photons \cite{jiang_optically_2023, meesala_non-classical_2024}.

The same suspension that protects the mechanical mode from radiating into the substrate, however, severs the dominant path for thermal anchoring. Under the optical pumping required for electro-optomechanical transduction, parasitic absorption heats the mechanical mode, and weak thermal anchoring lets that heat accumulate  \cite{meenehan_pulsed_2015, jiang_optically_2023}, forcing slower duty cycles or lower pump powers. Two-dimensional suspended designs relieve this trade-off only partially, and thermal noise remains a limitation \cite{sonar_high-efficiency_2025}.

\label{int:release_free}
A complementary route is to remove the suspension entirely and let the device layer remain in full contact with the substrate, trading the low radiation loss of a free-standing structure for a direct thermal path to the chip. The challenge is then to confine the mechanical mode without the protection of a full phononic band gap. We previously showed that this is possible by operating below the continuum of surface and bulk acoustic waves in the substrate, exploiting total internal reflection and geometrical softening to keep the mechanical mode confined within the device layer \cite{kolvik_clamped_2023, burger_design_2025}. The approach has been demonstrated for the silicon optomechanical crystal \cite{kolvik_clamped_2023, kolvik_optomechanical_2025}, and recently extended to a complete piezo-optomechanical transducer \cite{burger_design_2025, burger_release-free_2026} operating at room-temperature. The electromechanical crystal (EMC) -- the piezoelectric half of that transducer, responsible for the microwave-to-mechanics conversion -- has not yet been characterized in isolation, and whether the strong-coupling regime needed for high-fidelity quantum electromechanical operation can be reached in a release-free architecture has remained an open question.

\label{int:this_work}
Here, we demonstrate a release-free electromechanical crystal cavity in thin-film lithium niobate, integrated with a high-impedance NbTiN microwave resonator, that operates in the strong coupling regime. We measure EMCs on both silicon and sapphire substrates at room and cryogenic temperatures, reaching internal mechanical quality factors above $10^4$ at millikelvin temperature. Once integrated with the microwave resonator, the EMC reaches an electromechanical coupling rate $\gem/(2\pi) \approx \SI{30}{\mega\hertz}$ that exceeds both the mechanical and microwave loss rates. These results establish release-free EMCs as compact, scalable interfaces between microwaves and gigahertz sound, completing the electromechanical piece of the release-free piezo-optomechanical transducer architecture and opening a path toward low-noise microwave-to-optical quantum transduction.

\begin{figure*}[ht!]
    \centering
    \includegraphics[width=\textwidth]{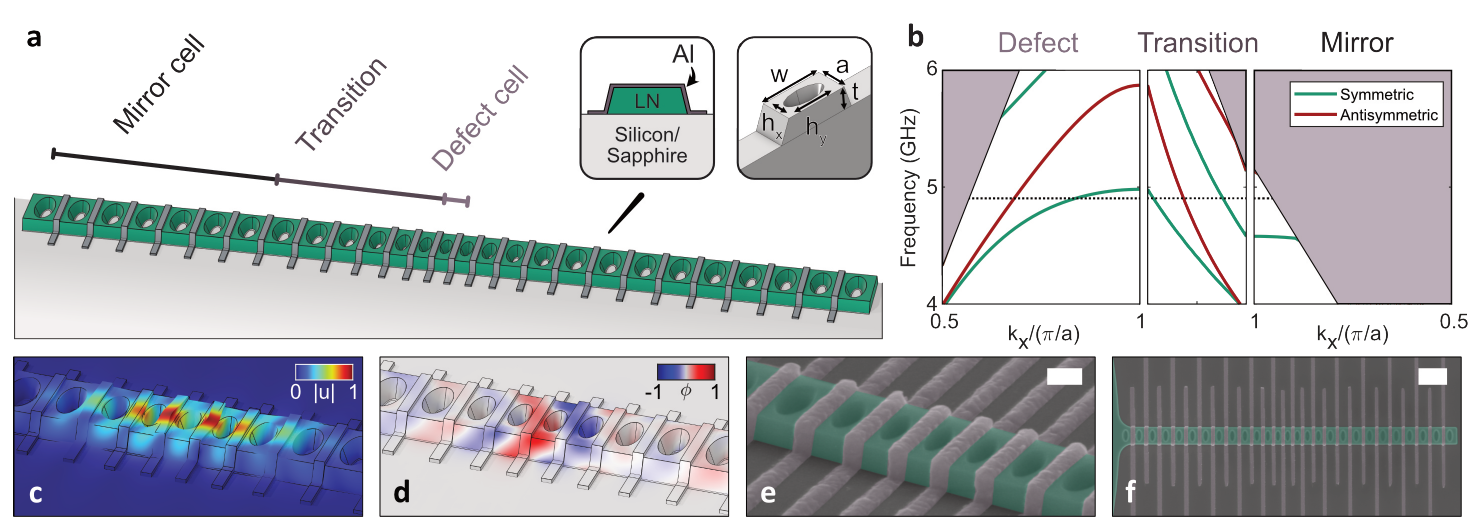} 
    \caption{
    \textbf{Release-free phononic crystal cavity with microwave readout.} 
    \textbf{a}, Schematic of the device design, showing the lithium niobate electromechanical crystal (green) and aluminum IDT (gray) going across with a defect cell in the center transitioning into mirror cells on the edges. Left inset: cross-sectional view of the device with materials indicated. Right inset: Unit cell geometry with the main design parameters labeled. \textbf{b}, Mechanical band diagram of the defect cell (left) and mirror cell (right), with the frequency of our main mechanical mode indicated by the gray dotted line for silicon substrate. The transition region (middle) shows the evolution of the X-point modes for a linear change in the unit-cell geometry from the defect cell to the mirror cell. \textbf{c}, Finite-element method simulation of our main mechanical mode, showing exaggerated deformation and normalized displacement magnitude $|u|$ in color. \textbf{d}, Normalized electrostatic potential $\phi$ for the mechanical mode in \textbf{b}.  \textbf{e}, False-color SEM (scanning electron microscope) image of the device, with materials colored as in \textbf{a} (scale bar \SI{200}{\nano\meter}). \textbf{f}, False-color SEM image of the complete electromechanical crystal viewed from above, with materials colored as in \textbf{a} (scale bar \SI{1}{\um}).}
    \label{fig:device}
\end{figure*}

\section{Design} 
\label{sec:design}

\label{design:waveguide confinement}
At the heart of our device lies a one-dimensional phononic crystal patterned in a 150-nm-thick film of X-cut lithium niobate on a silicon or sapphire substrate (Fig.~\ref{fig:device}a). We engineer a defect mode near the X-point of the band structure (Fig.~\ref{fig:device}d) with a pinch-type strain field whose dominant component $S_{xx}$ couples to the in-plane electric field of an interdigital transducer (IDT) patterned on top of the waveguide (Fig.~\ref{fig:device}c). The mode lies below the substrate sound cone defined by surface and bulk acoustic waves (SAW and BAW), so it cannot phase-match to any propagating substrate mode and is confined to the lithium niobate through total internal reflection. Lithium niobate has a lower bulk sound velocity than common substrates such as silicon and sapphire \cite{fu_phononic_2019, mayor_gigahertz_2021}, and the periodic patterning lowers the mode's phase velocity further through geometric softening \cite{burger_release-free_2026, sarabalis_release-free_2017, sarabalis_guided_2016,kolvik_clamped_2023} -- keeping the mode confined to the device layer even on acoustically slower substrates.

To localize the mode along the waveguide, we surround the defect cell with a mirror region supporting a quasi-bandgap at the mechanical resonance frequency (Fig.~\ref{fig:device}d), adiabatically tapered over several transition cells to suppress scattering into the substrate. The result is a release-free phononic cavity that, by analogy to the optomechanical crystal \cite{eichenfield_optomechanical_2009, chan_optimized_2012}, we call an electromechanical crystal.

\label{design:coupling}
The IDT couples the EMC mode either to a \SI{50}{\ohm} microwave feedline, at an external rate $\gammac$ (Sec.~\ref{sec:rt_meas} and~\ref{sec:cryos11}), or to a microwave resonator, at an electromechanical rate $\gem$ (Sec.~\ref{sec:nbtin_emc}). We obtain both from finite-element simulations through the piezoelectric overlap integral of Ref.~\cite{burger_design_2025}, and optimize the defect and mirror geometries with a Nelder--Mead algorithm to maximize $\Qi\gem$, with $\Qi$ the internal phononic quality factor. Across a wide range of initial conditions the optimization returns high-$\Qi$ modes, indicating that radiation loss does not limit the release-free architecture and that strong confinement is available throughout a large design space. We optimize for silicon and test the geometry with both lithium niobate on silicon (LNOS) and sapphire (LiSa) material stacks to evaluate transferability of the thin-film design.

\label{design:final design}
The main remaining lever is cavity length. A longer cavity raises the electrical participation and thus $\gem$, and permits a more gradual defect-to-mirror transition that lowers radiation loss, but it also lowers fabrication yield and crowds the mechanical spectrum (App.~\ref{app:len}). We therefore select a compact cavity of one defect, six transition, and seven mirror cells, with IDT electrodes covering all cells, prioritizing yield and device-to-device stability over peak coupling. We take the fundamental cavity mode as our mode of interest (Fig.~\ref{fig:device}a), for which we simulate an external feedline coupling rate $\gammac/(2\pi)$ up to \SI{67}{\kilo\hertz} -- equivalent to $\gem/(2\pi)=63\,\text{MHz}$ if coupled to a $\SI{10}{\femto\farad}$ microwave resonator -- and a radiation-limited $\Qi$ above the material-loss limit (Fig.~\ref{fig:device}e). The design choices are optimized for silicon while remaining compatible with other substrates.
 
\begin{figure*}    
    \centering
    \includegraphics[width=\linewidth]{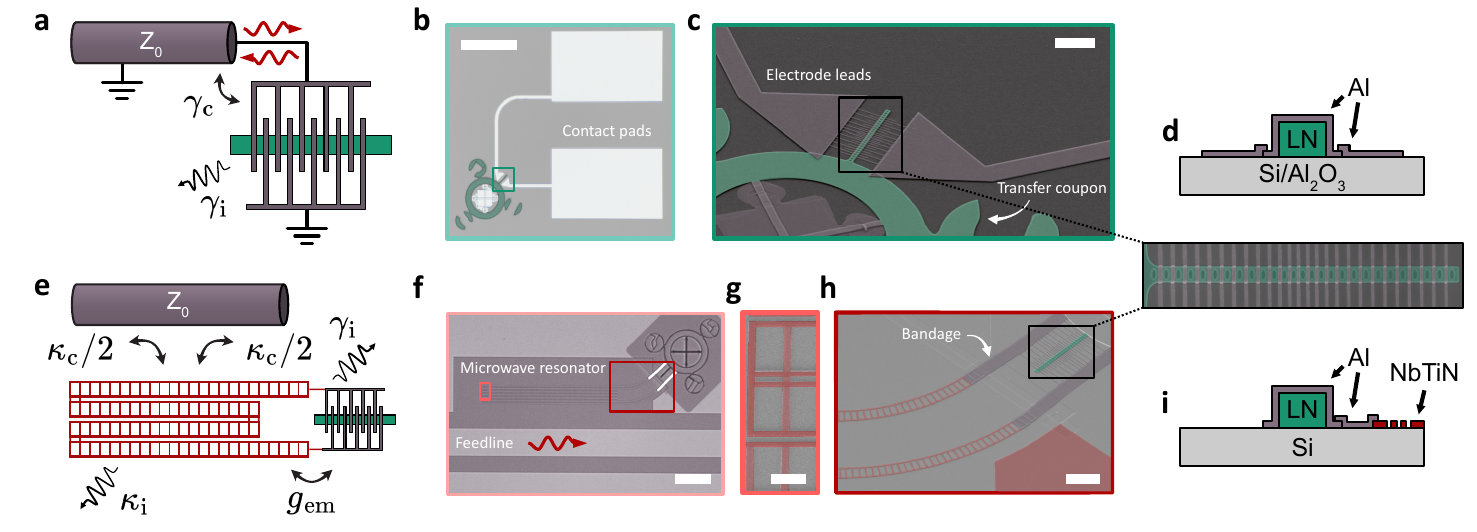}
    \caption{ \textbf{Fabricated release-free EMCs connected to microwave feedline (top) and high-kinetic-inductance microwave resonator (bottom).} Material color coding in schematics and false-color images denotes lithium niobate (green), NbTiN (red), and aluminum (dark gray). \textbf{a}, Independent EMC device configured for microwave reflection measurements. \textbf{b}, Micrograph of the EMC device with IDT connected to two probe pads (scale bar \SI{50}{\um}). \textbf{c}, False-color SEM of the electrode connections to to the EMC IDTs (scale bar \SI{5}{\um}). \textbf{d}, Cross section of bare EMC chip with materials indicated. \textbf{e}, Coupled system of an EMC and a microwave resonator, configured for microwave transmission measurement. \textbf{f}, Micrograph of the combined EMC and microwave resonator (scale bar \SI{25}{\um}). \textbf{g}, False-color SEM of small section of the microwave resonator (scale bar \SI{1}{\um}). \textbf{h}, False-color SEM of the connection between the microwave resonator and the EMC (scale bar \SI{5}{\um}). \textbf{i}, Cross section of coupled system with materials indicated. }
    \label{fig:devices}
\end{figure*}

\label{sec:devices}
\section{Device Implementations}
We implement the EMC in two architectures. In the first, the IDT connects directly to a microwave feedline via two aluminum contact pads (App.~\ref{app:fab_emc}), enabling electrical readout of the bare mechanical resonance on both silicon and sapphire substrates (Fig.~\ref{fig:device}a--d); this stand-alone device is the subject of Secs.~\ref{sec:rt_meas} and \ref{sec:cryos11}. In the second, the same EMC is integrated with a high-impedance microwave resonator to reach the strong coupling regime (Fig.~\ref{fig:device}e--i), as studied in Sec.~\ref{sec:nbtin_emc}.

The microwave resonator consists of a meandering NbTiN waveguide periodically patterned with square holes \cite{meesala_non-classical_2024}, connected to each end of the EMC IDT through aluminum bandages (App.~\ref{app:fab_nbtin_emc}). An external magnetic field threading the holes drives screening currents that modify the kinetic inductance, tuning the resonator frequency as $\Delta f_\mu / f_\mu \approx -k B_\mathrm{ext}^2$ \cite{xu_frequency-tunable_2019}. Its high kinetic inductance yields a characteristic impedance $Z_\mathrm{\mu,tot} = 2200\,\Omega$ in a \SI{133}{\um} footprint, with a low capacitance $C_\mathrm{\mu,tot} = \SI{13.4}{\femto\farad}$ (App.~\ref{app:s21model}).

\label{exp:nbtin:coupling}
We build the microwave resonator from niobium titanium nitride (NbTiN). Its short quasiparticle lifetime keeps the superconductivity resilient in optically intense environments \cite{lobo_photoinduced_2005, visser_quasiparticle_2014}, a prerequisite for the optical pumping needed in future microwave-to-optics transduction. NbTiN also offers a high sheet kinetic inductance, letting the meandering waveguide reach a characteristic impedance $Z_\mathrm{\mu,tot} = \sqrt{L_\mu/(\Cmu + \CIDT)}$ above \SI{1}{\kilo\ohm} (Fig.~\ref{fig:device}e). This benefits the electromechanical coupling: at fixed resonance frequency, a higher impedance means a lower capacitance $\Cmu$, raising the participation of the IDT capacitance as $\gem \propto \CIDT/(\Cmu + \CIDT)$.

\section{Room Temperature Characterization} 
\label{sec:rt_meas}

\begin{figure*}
    \centering
    \includegraphics[width=\textwidth]{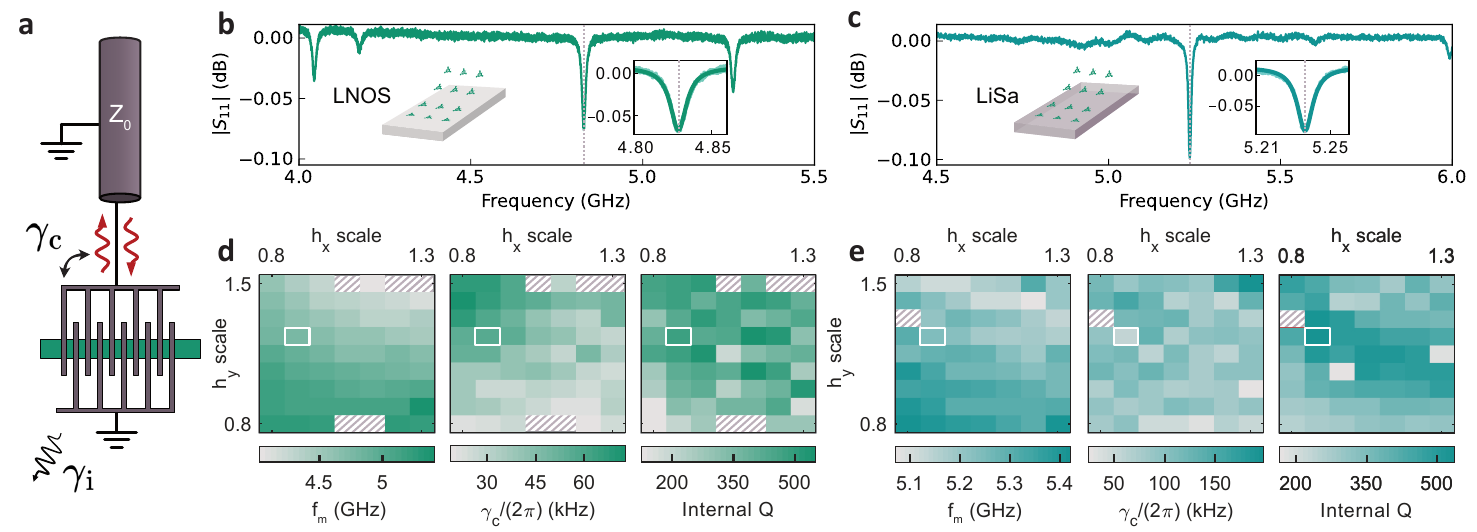}
    \caption{ \textbf{Room-temperature microwave measurement of release-free phononic crystal cavities on silicon and sapphire substrates.}
    \textbf{a}, Reflection measurement schematic showing the EMC coupled to a microwave feedline with impedance $Z_0$ through external coupling rate $\gammac$ and with internal loss rate $\gamma_i$. \textbf{b},(\textbf{c},) Microwave reflection spectrum $S_{11}$ of LNOS and LiSa EMCs measured at room temperature. Inset shows fit for indicated resonance. \textbf{d},(\textbf{e},) Measured sweep of ellipse sizes in the defect cell for LNOS (LiSa) EMCs. The resonance frequency $f_\mathrm{m}$ (left), coupling rate $\gammac$ (middle) and internal quality factor $\Qi$ (right) is shown for each device. White outline indicates device shown in \textbf{b}(\textbf{c}). Striped squares show non-working devices after fabrication.}
    \label{fig:RTs11}
\end{figure*}

\label{exp:RT_S11}
To characterize the EMCs at room temperature we use microwave probes to connect a vector network analyzer to the electrode pads (Fig. \ref{fig:RTs11}a). The devices are measured and analyzed using microwave reflection spectrum for chips on silicon (Fig. \ref{fig:RTs11}b,c) and sapphire (Fig. \ref{fig:RTs11}d,e). At room temperature our system is strongly undercoupled due to  the small IDT capacitance $\CIDT$ and large material losses, so we calibrate the measurement setup to remove background signals and enhance device visibility. We fit the  complex reflection signal $S_{11}$ to the equation \cite{wollack_loss_2021}
\begin{equation*}
    S_{11} = 1-\frac{2\gammac}{\gamma}\frac{e^{i\phi}}{1+2i(\omega-\omegam)/\gamma}
\end{equation*}
where $\phi$ accounts for impedance mismatches in the circuit, $\omega_m$ the mechanical resonance frequency, $\gammai$, $\gammac$ and $\gamma$ the internal, external and total mechanical loss rate, related by $\gamma = \gamma_i + \gammac$ (Fig. \ref{fig:RTs11}a). 

\label{exp:modes}
To characterize the EMCs we sweep the hole size of the defect cell, which also changes the transition region. Each EMC supports a small number of well-resolved modes within the measured frequency range, typically yielding four or fewer modes with strong responses (App. \ref{app:modes}). We track the dominant mode on each substrate (inset Fig. \ref{fig:RTs11}b-c), and fit the quality factor, coupling rate and frequency for different geometries (Fig. \ref{fig:RTs11}d-e).

\label{exp:rts11:fr}
Changes to the defect hole size also affect the transition cells, leading to geometry changes to the full cavity length. By changing the defect cell holes with a scaling factor between $0.8-1.5$, we measure a mechanical frequency covering a range of \SI{1}{\giga\hertz} for fundamental modes of the LNOS EMC, demonstrating a persevering mechanical resonance. Since the base design is chosen to be the same for both LNOS and LiSa, differences in behavior is largely due to substrate properties effect on the mechanical mode. In sapphire we observe modes at a higher frequency, owing to the increased stiffness leading to higher confinement compared to silicon. 

\label{exp:rts11:yc}
Variations in coupling rate are expected due to changes in confinement and mode shape. We measure coupling rates of $\gammac/(2\pi) =39.0\pm13.9 $ for LNOS EMCs and $\gammac/(2\pi) =96.3\pm30.5$ for LiSa EMCs, with maximum measured values of $\gamma_\mathrm{c,max}/(2\pi)=72.6\,\text{kHz}$ and $ \gamma_\mathrm{c,max}/(2\pi)=198\,\text{kHz}$ respectively (Fig. \ref{fig:RTs11}d-e). Here, uncertainties denote the standard deviation across measured devices. These results confirm the strong coupling we simulate in sec. \ref{sec:design} for LNOS devices, with LiSa devices demonstrating higher coupling despite using a design not optimized for the substrate.

\label{exp:rts11:qi}
From the sweep shown in Fig.~\ref{fig:RTs11}d,h we find mean internal quality factors for LNOS (LiSa) EMC $384\pm104\,(387\pm88)$, demonstrating comparable room temperature losses in the two substrate platforms. Across all measured devices, the highest internal quality factors we observe at room temperature are of $\Qi=810\,(\Qi=690)$ for LNOS (LiSa) EMCs. There is no clear dependence of the quality factor on geometry changes in the fundamental LNOS mode (Inset in Fig. \ref{fig:RTs11}b), suggesting that radiation loss is negligible compared to material loss. 

Based on previous reports for comparable systems, we expect room-temperature quality factors of $\Qi = 400-800$, although the precise value depends on device design and fabrication details (App. \ref{app:loss:rt}) \cite{sarabalis_s-band_2020, jiang_efficient_2020}. The measured loss rates therefore agree well with expected material loss, limiting further improvements at room temperature. To determine if radiation loss $\gamma_\mathrm{rad}$ is a limiting contribution we reduce the material losses by cooling down the devices in the next section.

\section{Cryogenic Characterization}
\label{sec:cryos11}
\label{cryo: measurement}
\begin{figure*}[ht]
    \centering
    \includegraphics[width=\textwidth]{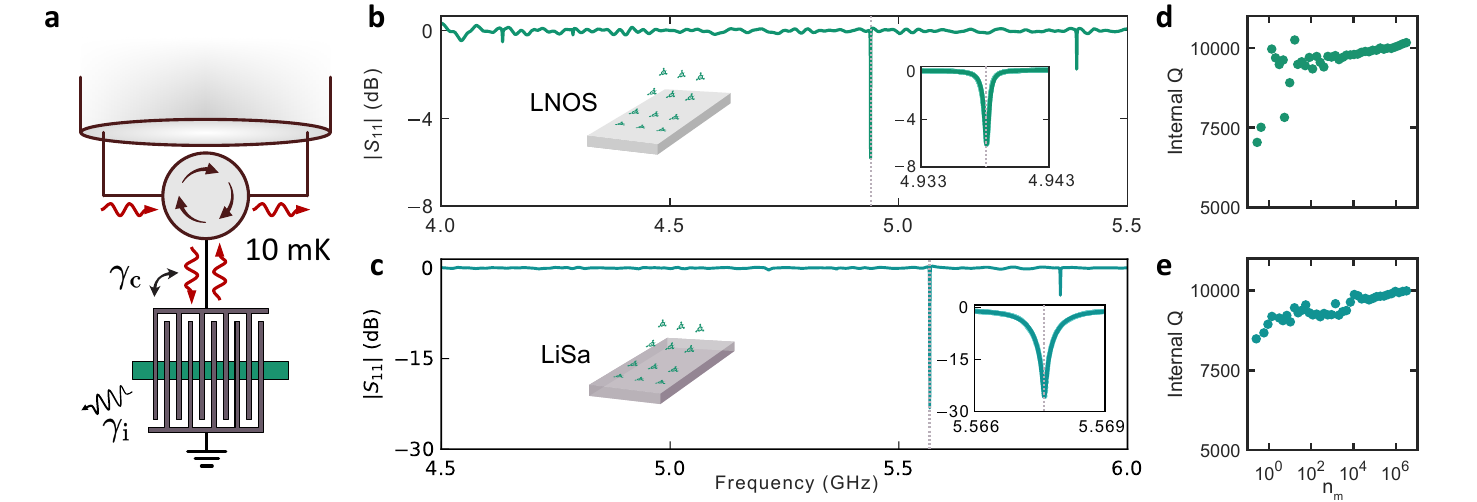}
    \caption{\textbf{Millikelvin microwave measurement of release-free phononic crystal cavities on silicon and sapphire substrates.} \textbf{a}, Cryogenic reflection measurement schematic showing the EMC connected to a circulator at the mixing chamber plate, with the external (internal) loss rate $\gammac$ ($\gamma_i$) indicated. \textbf{b},(\textbf{c},) Microwave reflection ($S_{11}$) measured for an LNOS (LiSa) EMC at cryogenic temperatures. Inset shows fit for center peak in \textbf{b}(\textbf{c}). For LiSa, the resonance shown corresponds to a different mechanical mode than the one tracked in Fig.~\ref{fig:RTs11}e. \textbf{d}, Internal quality factor as a function of intracavity phonon number $n_\mathrm{ph}$, shown for the LNOS and LiSa resonance shown in \textbf{b} and \textbf{c}, respectively.}
    \label{fig:cryoS11}
\end{figure*}

To investigate the mechanical loss mechanisms and characterize the EMC in the regime relevant for quantum application we cool down a select number of devices to \SI{10}{\milli\kelvin} inside a dilution refrigerator (App. \ref{app:cryo_setup}). To measure the EMC the electrode pads are wire bonded inside a printed circuit board and connected to a switch located at the mixing chamber. The signal is routed through a circulator, which allows us to measure the EMCs in reflection at different powers and fit as described previously (Fig. \ref{fig:cryoS11}a). 

\label{cryo:fr}
At cryogenic temperatures, we measure the microwave reflection spectrum and identify prominent mechanical resonances after digitally subtracting the broad background signal (Fig. \ref{fig:cryoS11}b-c). We find that the cold mechanical resonance frequencies are blue-shifted by 50-\SI{100}{\mega\hertz} compared to room temperature measurements. The frequency shifts are attributed to temperature dependent material stiffness \cite{shao_phononic_2019}. 

\label{cryo:qi}
The signal strength is also increased in part due to the increased internal quality factor at lower temperatures. During cool down internal material loss decrease as thermal phonons are frozen out and aluminum goes superconducting (App. \ref{app:loss:cryo}). Each mode will then exhibit an increased internal quality factor until it is limited by either radiation or material loss. We identify a number of modes showing significantly lower loss, and focus on the best performing mechanical mode for LNOS and LiSa EMC respectively.

\label{cryo:power}
We characterize the EMCs at different powers and find both LiSa and LNOS EMC reaching $\Qi = 10^4$ at high average phonon occupation $n_\mathrm{ph}$ (Fig. \ref{fig:cryoS11}d-e). At low phonon occupation we observe time-dependent fluctuations in the resonator response that limit accuracy of the model fit at low powers (App. \ref{app:loss}). We attribute this behavior, together with the reduced intrinsic quality factor at low $n_\mathrm{ph}$ to interactions with two-level systems which are expected to increase at low power \cite{wollack_loss_2021, maksymowych_spectral_2025}. 

From literature we expect material loss to limit internal quality factor to $Q_\mathrm{mat}\approx5\cdot 10^4$ at low phonon numbers and $Q_\mathrm{mat} \approx 2\cdot10^5$ at high phonon numbers (App. \ref{app:loss}). Our devices are therefore likely limited by radiation loss, either from design or fabrication disorder. From simulations we expect that design is not a limiting factor with radiation-limited quality factors up to $10^5$ and achieving high $\Qi$ is therefore expected to be mainly a fabrication challenge, although further design optimization remains for robustness against disorder. 

\label{cryo:gamma_c}
We measure a higher effective coupling $\gammac$ in the cryogenic setup compared to the room temperature coupling rate, with $\gammac/(2\pi)$ increasing from \SI{54}{\kilo\hertz} (\SI{119}{\kilo\hertz}) to \SI{117}{\kilo\hertz} (\SI{509}{\kilo\hertz}) for the LNOS (LiSa) devices in Fig. \ref{fig:cryoS11}. Resonances are only measurable below a temperature threshold at which internal losses are sufficiently reduced (App.~\ref{app:Tsweep}). In this regime, $\gammac$ exhibits only a weak temperature dependence, suggesting that the observed change in coupling rate does not arise from a sharp transition (e.g. the onset of superconductivity). Instead, we attribute the increased coupling to a difference in the microwave environment between the dilution refrigerator compared to the room-temperature measurement setup, which are not calibrated and may include contributions from surrounding structures. The increased ratio $\gammac/\gamma$ results in a strong signal, enabling measurement and operation close to critical coupling.

\section{Strong coupling between EMC and microwave resonator}
\label{sec:nbtin_emc}

\label{exp:nbtin:motivation}
Having characterized the bare EMC, we now integrate it with the high-impedance microwave resonator and ask whether the strong-coupling regime -- the prerequisite for coherent, high-fidelity quantum electromechanical operation -- can be reached in a release-free architecture. Beyond this specific question, the experiment tests whether release-free phononic crystal devices can be embedded in a realistic superconducting microwave circuit, a necessary step toward microwave-to-optics quantum transduction.

\begin{figure}[ht]
    \centering
    \includegraphics[width=\linewidth]{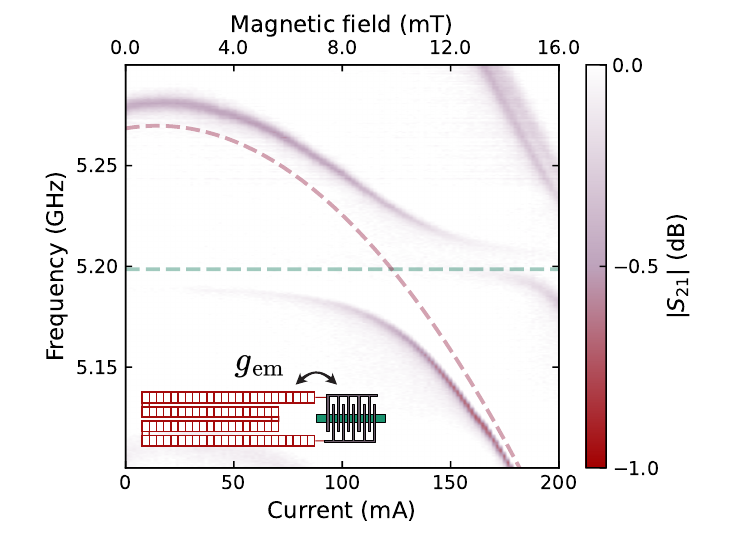}
    \caption{\textbf{Release-free electromechanical crystal in the strong-coupling regime with a high-impedance microwave resonator.} Microwave transmission $|S_{21}|$ measured at \SI{10}{\milli\kelvin} as the resonator is tuned through the mechanical mode with the applied coil current. Dotted lines mark the simulated bare microwave (red) and mechanical (green) resonance frequencies; where they intersect, the transmission splits into an avoided crossing of magnitude $2\gem/(2\pi) = \SI{60}{\mega\hertz}$, the hallmark of coherent electromechanical coupling. Inset: schematic of the coupled superconducting-electromechanical system.}
    \label{fig:nbtin_emc}
\end{figure}

\label{exp:nbtin:measurement}
We cool the device to \SI{10}{\milli\kelvin} in a dilution refrigerator (App.~\ref{app:cryo_setup}) and record the microwave transmission $S_{21}$ while tuning the resonator with an external superconducting coil (Fig.~\ref{fig:nbtin_emc}). A current of \SI{200}{\milli\ampere}, corresponding to an on-chip field $B_\mathrm{ext} = \SI{16}{\milli\tesla}$, sweeps the bare resonator frequency over \SI{200}{\mega\hertz} -- enough to bring it through the mechanical resonance. After removing the background and the electrical-delay phase, we fit the transmission at every current to (App.~\ref{app:s21model})
\begin{equation*}
    S_{21}(\omega) = 1 - \frac{|\kappac|}{\kappa + 2i(\omega-\omegamu) + \dfrac{2\gem^2}{\gamma + 2i(\omega - \omegam)}},
\end{equation*}
with $\kappa$ ($\kappac$) the total (external) microwave loss rate, $\gamma$ the total mechanical loss rate, and $\omegamu$, $\omegam$ the microwave and mechanical resonance frequencies.

\label{exp:nbtin:result}
As the resonator is tuned through the mechanical mode, the transmission reveals a clear avoided crossing (Fig.~\ref{fig:nbtin_emc}), the signature of coherent energy exchange between the two systems. From the splitting we extract an electromechanical coupling rate $\gem/(2\pi) = \SI{30}{\mega\hertz}$, together with quality factors $Q_{\mu,\mathrm{i}} = 1200$ ($Q_{\mu}=1100$) and $\Qm = 1200$ for the microwave and mechanical modes. The corresponding loss rates lie well below the coupling strength, placing the device firmly in the strong-coupling regime with a cooperativity $\mathcal{C}_\mathrm{em} = 4\gem^2/(\kappamu\gamma_\mathrm{m}) = 180$. To our knowledge this is the first release-free crystal device to reach this regime, demonstrating that removing the suspension need not come at the cost of coherent electromechanical coupling.

\label{exp:nbtin:lossmechanism}
Integration does, however, load the resonator: its internal quality factor falls to $Q_{\mu,\mathrm{i}} = 1200$, from $Q_{\mu,\mathrm{i}} = 75000$ for bare high-impedance resonators on the same chip. We hypothesize several contributions: a background of parasitic EMC modes that each open a decay channel into the resonator (App.~\ref{app:nbtin}), a Purcell-like decay $\kappa_{\mu,\mathrm{p}} = \gamma_m\,\gem^2/\Delta^2$ induced by the strong coupling to the main mechanical mode, and a more disordered superconducting film from the extended fabrication process. The last raises microwave loss but also the kinetic inductance, and with it the impedance and coupling, so these mechanisms set a trade-off between strong coupling and microwave coherence.

\label{exp:nbtin:outlook}
Much of this loss is avoidable. The full electrode coverage adds about \SI{2}{\femto\farad} of capacitance beyond the \SI{1.3}{\femto\farad} of the mechanically active region, raising microwave loss without improving $\gem$; trimming the electrodes should remove most of it while leaving the coupling intact (App.~\ref{app:len}). The coupling is in any case sufficient -- scaling from the highest external coupling rates $\gammac/(2\pi)=198\,\text{kHz}$ measured on sapphire projects $\gem/(2\pi)$ above \SI{100}{\mega\hertz} for this architecture -- so the device is limited by microwave and mechanical losses rather than by $\gem$, making $Q_\mu$ and $Q_m$ the key targets for future improvement.


\section{Discussion}
\label{sec:discussion}
\label{dis:core_result}
We have demonstrated a release-free electromechanical crystal that reaches the strong-coupling regime with a high-impedance microwave resonator, its coupling rate exceeding both the mechanical and microwave loss rates. The silicon-optimized geometry transfers to sapphire with comparable coupling and quality factors, and a wide sweep of defect geometries consistently yields a well-confined mode -- so release-free phononic crystals remain compatible with state-transfer operations between microwave photons and phonons.

\label{dis:positioning}
The central appeal of the platform is thermal: full contact with the substrate opens a direct path for optically generated heat to escape into the chip. In suspended optomechanical crystals this heat accumulates in the mechanical mode, forcing the short duty cycles and reduced pump powers that bottleneck microwave-to-optical transduction \cite{meenehan_pulsed_2015, jiang_optically_2023}; release-free devices are designed to relax exactly this constraint \cite{kolvik_optomechanical_2025, burger_release-free_2026}, and the strong electromechanical coupling shown here supplies the piezoelectric half of such a transducer \cite{burger_design_2025}. Placement on the substrate also removes the sacrificial release step and frees the choice of substrate -- high-resistivity silicon or sapphire for low-loss microwave circuitry -- which, together with the wavelength-scale footprint and direct piezoelectric readout, makes the EMC a natural building block for heterogeneous integration with photonic, phononic, and superconducting systems \cite{yamagata_surface-acoustic-wave_2023, yao_perspectives_2025, zhang_scalable_2025}. The design extends below the gigahertz band as well, where larger devices with higher IDT capacitance and the favorable low-frequency losses of aluminum and lithium niobate \cite{schaffer_measurement_2023, shao_phononic_2019} open a route to critical coupling at room temperature for classical filtering, sensing, and modulation.

\label{dis:rtvscryo}
The present devices are limited by imperfections rather than design. Internal quality factors reach $\Qi \sim 810$ at room temperature, consistent with material loss, and $10^4$ at millikelvin temperature, where the weak power dependence points to radiation loss; since simulations place the radiation-limited $\Qi$ well above both, the gap is most likely fabrication disorder rather than the architecture itself. Integration with the superconducting circuit enhances the coupling as intended but adds microwave resonator loss from the EMC, and disentangling these channels would both improve performance and guide the EMC's use in other hybrid systems.

\label{dis:outlook}
The path forward is thus clear: higher quality factors, less parasitic capacitance, and lower device-to-device variability should follow from refinements to the design and fabrication, and the coupling budget already permits it, with rates above \SI{100}{\mega\hertz} projected for this architecture. These results establish release-free phononic crystals as a viable, flexible platform for integrated piezomechanical systems, and lay the groundwork for the compact, thermally anchored electromechanical devices needed for microwave-to-optical quantum transduction.

\noindent\textbf{Funding.}$\ $\normalsize
We gratefully acknowledge support from the Wallenberg Centre for Quantum Technology and from the European Research Council via Starting Grant 948265.
\newline

\noindent\textbf{Acknowledgement.}$\ $\normalsize
We acknowledge David Hambraeus, Witlef Wieczorek, Nils Johan Engelsen, and Per Delsing for helpful discussions. We thank Niclas Lindvall for his assistance in developing a custom alignment method and Vittorio Buccheri for his guidance on NbTiN fabrication. The devices were fabricated in the MyFab Nanofabrication Laboratory at Chalmers.
\newline

\noindent\textbf{Data availability.}$\ $\normalsize
The datasets generated and analyzed for the current study are available from the corresponding author on reasonable request.
\newline

\bibliography{MyLibrary} 

\appendix
\section{EMC Fabrication}
\label{app:fab_emc}
\begin{figure*}[ht!]
\centering
\includegraphics[width=\textwidth]{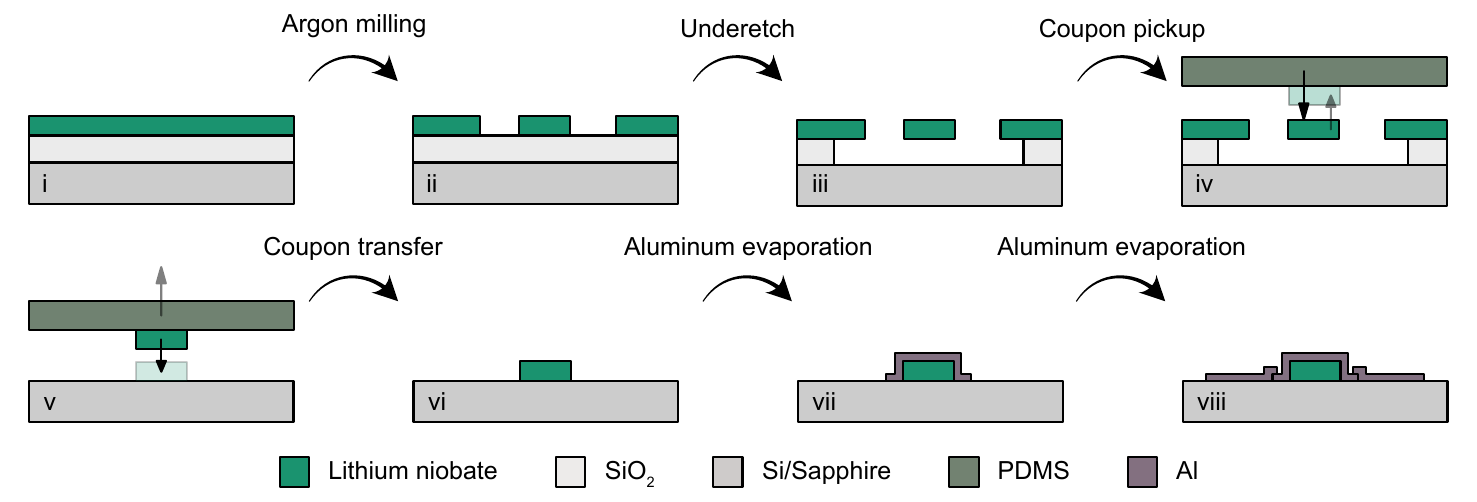}
\caption{\textbf{Fabrication process flow for EMCs on either a silicon or a sapphire substrate. } \textbf{i},  Blanket ion milling of \SI{300}{\nm} lithium niobate to a thickness of \SI{150}{\nm} on an oxide substrate. \textbf{ii}, Patterning of the coupon and EMC structures by argon milling. \textbf{iii}, Suspension of the coupon by wet etching. \textbf{iv}, Pick-up of the coupon using a PDMS stamp. \textbf{v}, Transfer of the coupon onto the target chip. \textbf{vi}, Cleaning and annealing of the micro-transferred chip. \textbf{vii}, Angled deposition of aluminum IDT electrodes. \textbf{viii}, Deposition of aluminum electrodes and pads.}
\label{fig:emc_fab}
\end{figure*}

The following steps describe the fabrication process for the LNOS EMC, and the devices presented in Sections \ref{sec:rt_meas} and \ref{sec:cryos11} (Fig. \ref{fig:emc_fab}). Electron-beam lithography (Raith EBPG 5200) is used for all lithography steps. For devices fabricated on sapphire substrates, a thin chromium layer is additionally deposited on top of the resist during all electron-beam lithography steps to mitigate charging effects and provide a height reference for focus and alignment. 

A chip consisting of 300 nm X-cut lithium niobate on \SI{4.7}{\um} of silicon oxide (NanoLN) is first thinned to \SI{150}{\nano\meter} by ion beam etching (Oxford Ionfab 300+). The EMC structures and release coupons are patterned in a negative electron-beam resist (FOx-16:MIBK  10:3 dilution) using electron-beam lithography and transferred into the lithium niobate layer by ion beam etching. 

The resist is removed using BOE, followed by cleaning in piranha solution and a 2\% HF dip. The coupons are subsequently released from the underlying \ce{SiO2} layer using BOE for 17 min.  

The suspended coupons are transferred to the target substrate by micro-transfer printing. The coupons are picked up using a polydimethylsiloxane (PDMS) stamp fabricated from Sylgard 184 and printed onto the receiving substrate, either silicon ([100] Silicon Siegert Wafers) or sapphire (C-plane Sapphire Siegert Wafers). To improve adhesion between the lithium niobate coupons and the receiving substrate, the samples are annealed at \SI{500}{\celsius} in an air atmosphere. 

Following a second piranha clean and HF dip, the IDT electrodes are patterned in AR-P 6200.09 electron-beam resist (CSAR 62, Allresist GmbH) using electron-beam lithography (Raith EBPG 5200). Local translational and rotational offsets introduced during the micro-transfer-printing process are corrected using a custom marker-detection and alignment procedure (App. \ref{app:coup}). The IDT electrodes are deposited by angled aluminum evaporation (Plassys) at ($-66^\circ$), ($0^\circ$), and ($66^\circ$) to achieve a uniform metal thickness on both the waveguide sidewalls and top surface. The resist is subsequently removed by lift-off in a hot NMP bath.

Large contact pads are then patterned in AR-P 6200.13 electron-beam resist (CSAR 62, Allresist GmbH) using electron-beam lithography. Prior to metal deposition, a breakthrough etch is performed to remove the native oxide and ensure electrical contact to the underlying electrode layer. Aluminum is evaporated normal to the substrate surface, followed by lift-off in a hot NMP bath.

\section{NbTiN \& EMC Fabrication}
\label{app:fab_nbtin_emc}
\label{fab:lnos_nbtin}
The following section describes the fabrication process used to realize the integrated superconducting-mechanical system presented in Section \ref{sec:nbtin_emc} (Fig. \ref{fig:emc_nbtin_fab}). Fabrication follows the process described in Appendix \ref{app:fab_emc} up to the transfer printing of the lithium niobate coupons, which are printed onto a silicon chip. Following transfer printing, excess lithium niobate is removed by electron-beam lithography and ion-beam etching, and local alignment markers are defined for each EMC device. Prior to subsequent processing, the surface is cleaned in piranha solution followed by a 2\% HF dip. For all subsequent lithography steps, resist removal and metal lift-off are performed using a hot NMP bath.

\begin{figure*}[ht!]
\centering
\includegraphics[width=\textwidth]{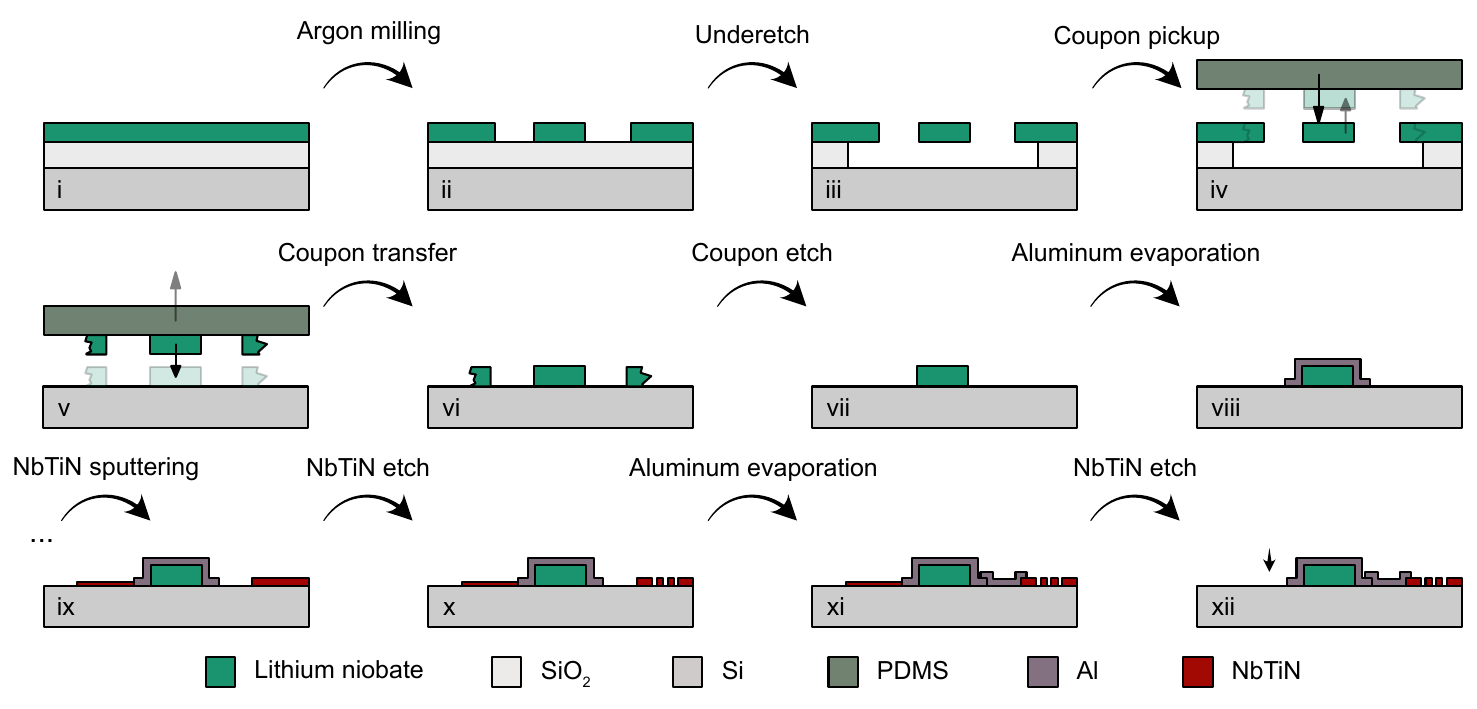}
\caption{\textbf{Fabrication process flow for EMCs integrated with a high-impedance microwave resonator.} \textbf{i}, Blanket ion milling of \SI{300}{\nm} lithium niobate to a thickness of \SI{150}{\nm} on an oxide substrate. \textbf{ii}, Patterning of the coupon and EMC structures by argon milling. \textbf{iii}, Suspension of the coupon by wet etching. \textbf{iv}, Pick-up of the coupon using a PDMS stamp. \textbf{v}, Transfer of the coupon onto the target chip. \textbf{vi}, Cleaning and annealing of the micro-transferred chip. \textbf{vii}, Removal of the lithium niobate coupon by argon milling. \textbf{viii}, Angled deposition of aluminum IDT electrodes. \textbf{ix}, Sputter deposition of a \SI{14}{\nm} NbTiN layer. \textbf{x}, Patterning of the microwave circuit by reactive ion etching. \textbf{xi}, Deposition of aluminum bandages and contact pads. \textbf{xii}, Removal of the short-circuiting NbTiN strip by argon milling.}
\label{fig:emc_nbtin_fab}

\end{figure*}

The IDT electrodes are first patterned in AR-P 6200.09 electron-beam resist and deposited by three-angle aluminum evaporation as described in Appendix \ref{app:fab_emc}, followed by lift-off. The EMC structures are subsequently protected using AR-P 6200.09 resist, and a \SI{14}{\nano\meter}-thick NbTiN film is deposited by reactive sputtering of a NbTi target in an argon/\ce{N2} atmosphere (DCA Instruments cluster system). Lift-off is then used to prevent NbTiN deposition on the EMC structures while retaining NbTiN in the surrounding regions.

The microwave circuitry is patterned in AR-P 6200.13 resist and transferred to the NbTiN film by reactive ion etching in an \ce{Ar}/\ce{HCl} plasma (Oxford Plasmalab 100). This approach is used because the narrow features of the high-impedance resonator are not reliably defined by direct NbTiN lift-off. Following resist removal and cleaning, electrical connections between the NbTiN circuitry and the aluminum IDT electrodes are formed using a second electron-beam lithography step in AR-P 6200.13 resist. A breakthrough etch is performed to remove the native oxide from both metal layers, after which aluminum is evaporated (Plassys) to form interconnect bandages. The resist is subsequently removed by lift-off.

Finally, a lithography and etch step using AR-P 6200.13 resist is performed to prevent short-circuit from the NbTiN fence formed along the edge of the sputter mask. This residual NbTiN is sufficiently thick to electrically short the device but cannot be removed during the initial circuit etch without introducing significant lateral etching of the patterned microwave circuitry. The completed devices are then cleaned and prepared for wire bonding.

\section{Coupon Alignment}
\label{app:coup}
\label{app:coup:transfer}
During micro-transfer printing, the entire array of devices is transferred in a single step using coarse alignment to the target chip. The transfer process introduces local displacement and rotation of individual coupons relative to the global chip coordinate system, likely due to stretching and deformation of the PDMS stamp. Subsequent fabrication steps, however, require alignment accuracies on the order of \SI{20}{\nano\meter} for electron-beam lithography of the IDT electrodes. The resulting local displacement of the EMC coupons therefore prevents the use of conventional global alignment procedures.

\begin{figure}[ht!]
    \centering
    \includegraphics[width=\linewidth]{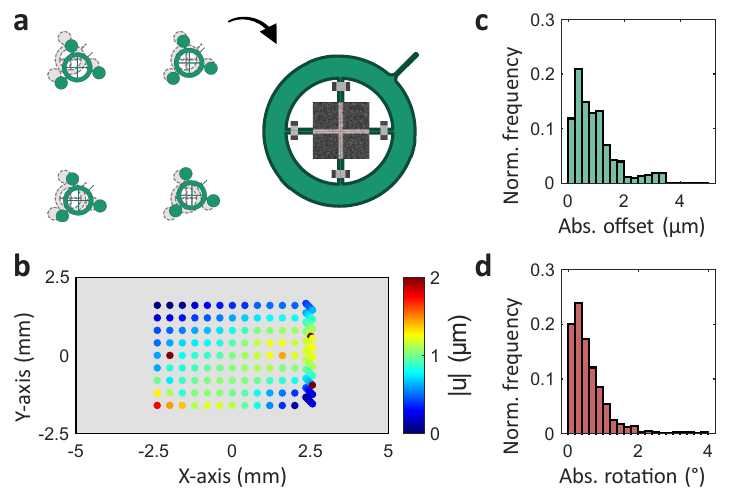}
    \caption{\textbf{Local displacement and rotation of devices during micro-transfer printing.} \textbf{a}, Schematic illustrating local offset of each lithium niobate coupon after micro-transfer printing, with local detection of one coupon illustrated. \textbf{b}, Scatter plot with each coupon location shown and their absolute offset shown in color. Missing or faulty detections are set to have max amplitude. \textbf{c}, Histogram plot showing absolute displacement of coupon on one chip. \textbf{d}, Histogram plot showing rotation of coupons on one chip.}
    \label{fig:Offset}
\end{figure}

\label{app:coup:align}
To achieve local alignment, both the displacement and rotation of each coupon must be determined from a single alignment marker, a capability not supported by the standard alignment routines in our electron-beam lithography software. We therefore employ a custom alignment routine that extracts both quantities from the central cross marker on each lithium niobate coupon while minimizing undesired resist exposure near the EMC due to proximity effects (Fig.~\ref{fig:Offset}a).

\label{app:coup:marker}
The detected marker position is used to determine the local displacement relative to the global coordinate system as well as the local rotation of each coupon (Fig.~\ref{fig:Offset}b--d). For fabrication layers with relaxed alignment requirements, the measured displacement is incorporated into the pattern design, allowing these layers to be exposed using global alignment despite the local offset. Besides being required for the microwave circuit etch layer (Fig.~\ref{fig:emc_nbtin_fab}xiii), this approach substantially reduces total exposure time and is therefore advantageous whenever the alignment tolerance permits.

\label{app:coup:displacement}
The transferred devices exhibit local displacements of up to \SI{3}{\um} and rotations of up to $2^\circ$. Even small rotational errors $\delta\theta$ can produce substantial positional errors at the EMC location, as the displacement scales with the distance $r$ from the alignment marker according to $\delta l = r\delta\theta$. For a rotation of $2^\circ$, the resulting misalignment at the EMC is approximately \SI{1.3}{\um}, comparable to the measured translational offsets.

\label{app:coup:spatial}
Mapping the local displacement across the chip reveals a strong spatial dependence (Fig.~\ref{fig:Offset}b), consistent with stretching and bending of the PDMS stamp during contact with the substrate in the micro-transfer printing process.

\section{EMC Modes}
\label{app:modes}
\begin{figure}[ht!]
    \centering
    \includegraphics[width=\linewidth]{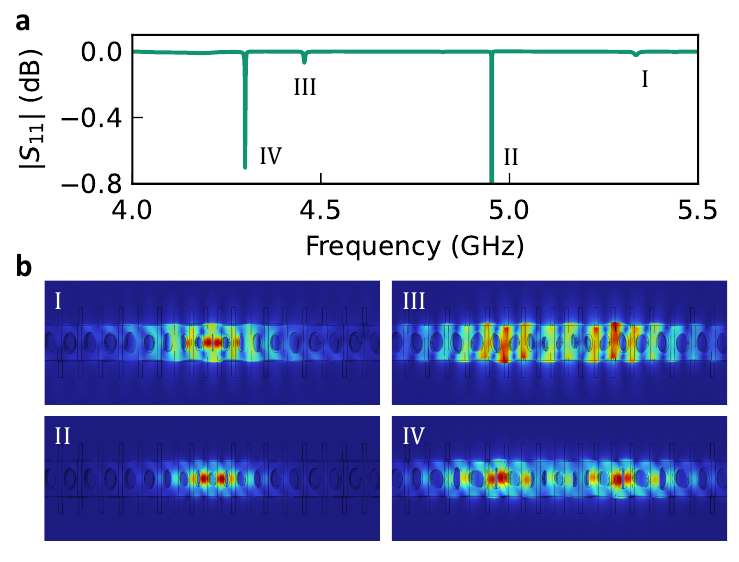}
    \caption{\textbf{Modes with strong microwave reflection signal in an optimized LNOS EMC design.} \textbf{a}, Simulated microwave reflection signal of the electromechanical crystal, optimized for low loss and high coupling of mode (II). \textbf{b}, Mechanical mode shape of the four most prominent resonance peaks indicated in (\textbf{a}).}
    \label{fig:modes}
\end{figure}

We observe several mechanical modes in the microwave reflection response, exhibiting different external coupling rates and intrinsic quality factors. To identify the dominant resonances, we simulate the full frequency response of an optimized LNOS electromechanical crystal (EMC) device and extract the modes with the strongest microwave response (Fig. \ref{fig:modes}a-b). The simulated spectrum reproduces the main features observed experimentally in Section \ref{sec:rt_meas}-\ref{sec:cryos11}.

\label{Modes}

Modes I and II originate from the two lowest-frequency symmetric mode branches identified in the unit-cell simulations. Mode II corresponds to the pinch mode shown in Fig. \ref{fig:device} and is the mode targeted by the device optimization. Mode I appears as a hybridized mode involving the pinch branch and a higher-frequency Rayleigh-like branch. While the exact character of Mode I varies somewhat between geometries, it consistently appears as a mixed symmetric mode.

Modes III and IV correspond to higher-order shear- and pinch-like modes. Although these are higher-order localized states, they appear at lower frequencies than the pinch mode due to the effective defect potential formed by the transition between the defect and mirror unit cells. These modes hybridize and their relative ordering in frequency depends on the device geometry, causing either mode to appear at the higher frequency. Nevertheless, the overall mode character remains consistent across the investigated designs.

The fundamental shear mode is also present in the simulated spectrum but exhibits only a weak microwave response due to its low electromechanical coupling. More generally, fundamental modes tend to exhibit lower coupling rates because of their smaller mode size, although they may also suffer from reduced intrinsic quality factors due to weaker confinement.

Additional mechanical modes exist within the structure but generally do not produce a strong microwave reflection signal, either because of weak coupling or low intrinsic quality factors. Depending on the geometry, some of these modes may hybridize with neighboring branches or exhibit enhanced coupling, leading to stronger microwave signatures.

\section{Frequency-Domain Simulation of the High-Impedance Microwave Resonator}
\label{app:nbtin}
\label{app:nbtin:coupling}
\begin{figure}
    \centering
    \includegraphics[width=\linewidth]{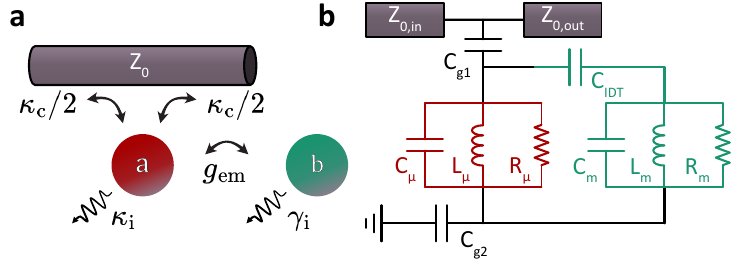}
    \caption{\textbf{Equivalent circuit model for coupled mode system measured in transmission.} \textbf{a}, Integrated EMC mode schematic indicating microwave mode ($\hat{a}$), mechanical mode ($\hat{b}$), their internal loss rates ($\kappai$) and ($\gammai$) respectively. The microwave mode is coupled to a feedline with coupling rate $\kappac$ and the the two modes interact with coupling rate $\gem$. \textbf{b}, Circuit representation of the device, indicating microwave resonator (red) and EMC mode with IDT (green). The microwave mode is then coupled capacitively coupled to the feedline and ground with $C_\mathrm{g1}$ and $C_\mathrm{g2}$ respectively. }
    \label{fig:circuit}
\end{figure}

\begin{figure*}[ht!]
    \centering
    \includegraphics[width=\linewidth]{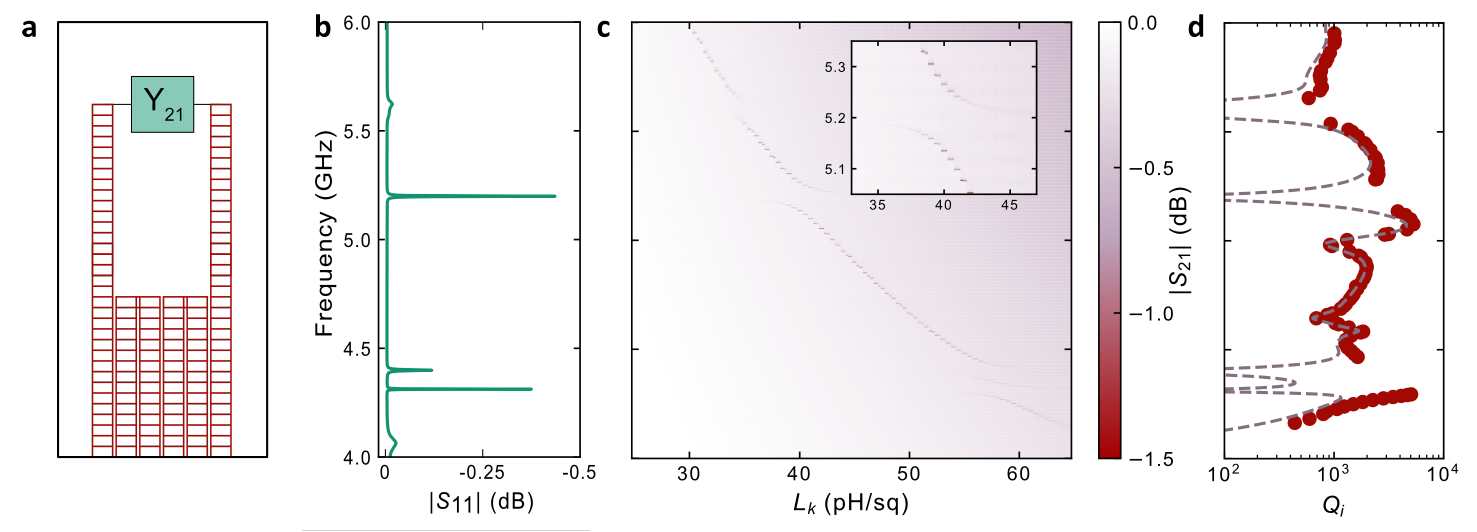}
    \caption{\textbf{Frequency-domain simulation of NbTiN-EMC system }\textbf{a}, Schematic illustrating how the EMC admittance is connected to simulate the full system response. Inset showing an avoided crossing with the main EMC mode at \SI{5.2}{\GHz}. \textbf{b}, Simulated one-port microwave reflection response $|S_{11}|$ of a silicon-clamped EMC. \textbf{c}, Color map showing the simulated microwave transmission response $|S_{21}|$ for different sheet kinetic inductance $L_k$ in the model. \textbf{d}, Fitted internal quality factor of the simulated response from \textbf{c} (red markers) with modeled internal quality factors modeled with the EMC admittance (dotted gray line).}
    \label{fig:nbtinsim}
\end{figure*}

To further analyze the electromechanical system (Fig. \ref{fig:circuit}a), we simulate the NbTiN microwave resonator in SONNET. We focus on the fundamental mode of an open-open meandering resonator, which creates voltage anti-nodes at each end for efficient electromechanical coupling. SONNET allows arbitrary admittances to be connected to selected regions of the metal layout (Fig. \ref{fig:nbtinsim}a), which we use to extract equivalent-circuit parameters of the microwave resonator and to analyze the coupled electromechanical device (Fig. \ref{fig:circuit}b). 

\label{app:nbtin:impedance}
By connecting a variable load capacitance $C_\delta$ as a probe admittance, we extract the lumped element parameters of the resonator. Since the resonance frequency depends on the total capacitance, the frequency shift induced by the load capacitance can be used to determine the bare resonator capacitance $C_\mathrm{\mu}$

 \begin{equation*}
    C_\mu = C_\delta\frac{\omega_{\mu,\delta}^2}{\omegamu^2-\omega_\mathrm{\mu,\delta}^2}
\end{equation*}

with $\omega_\mathrm{\mu}$ the bare microwave frequency and $\omega_\mathrm{\mu,\delta}$ the shifted resonance frequency. Using this method, we obtain a bare resonator capacitance $C_\mathrm{\mu}=10.1\,\text{fF}$, corresponding to a resonator impedance of $Z_\mathrm{\mu}= (\omega_\mathrm{\mu}C_\mathrm{\mu})^{-1}= 2.5 \,\mathrm{k\Omega}$. Including the IDT capacitance increases the total capacitance to $C_\mathrm{\mu,tot} = 13.4 \,\text{fF}$ and reduces the impedance to $Z_\mathrm{\mu,tot}= 2.2 \,\mathrm{k\Omega}$. 

After including the IDT capacitance the simulated resonance frequency is predicted to be $f_\mathrm{\mu} = 5.44\,\text{GHz}$. Although the resonator is designed to lie above the final device frequency to accommodate the frequency downshift introduced by the EMC, the predicted value remains $3\%$ higher than the measured device resonance. We attribute this discrepancy to parasitic capacitance and fabrication-induced variations in the sheet kinetic inductance from film thickness, nanowire width or deposition disorder. 

\label{app:nbtin:inductance}
With the capacitance determined, we next extract the relative contributions of kinetic and geometric inductance. The resonance frequency is determined by the two forms of inductance as 
\begin{equation*}
    \omega_\mathrm{\mu} = \frac{1}{\sqrt{C_\mathrm{\mu}(L_\mathrm{g}+\sigma_\mathrm{g}L_\mathrm{k,s})}}
\end{equation*}

where we express the total kinetic inductance as $L_\mathrm{k}=\sigma_\mathrm{g}L_\mathrm{k,s}$, with $L_\mathrm{k,s}$ the sheet kinetic inductance and $\sigma_\mathrm{g}$ a geometry factor. By sweeping the sheet kinetic inductance in the simulation and tracking the resulting resonance frequency shift, we extract a kinetic inductance of \SI{64.0}{\nano\henry} with a negligible geometric contribution of \SI{0.2}{\nano\henry}.

\label{app:nbtin:simulation}
To build a better understanding of our electromechanical system, we connect the admittance of the EMC as a load in the simulation model. The EMC admittance is extracted from a COMSOL model, where the geometry was adapted to include fabrication disorder and cryogenic material loss (Fig. \ref{fig:nbtinsim}a). To add in material loss we add an imaginary component to the density of the lithium niobate in our model. The admittance is connected to each lead of the NbTiN microwave resonator in our model, and the response is simulated for the full system (Fig. \ref{fig:nbtinsim}b). To sweep the resonance frequency of the microwave resonator, we again change the sheet kinetic inductance $L_\mathrm{k}$ of the NbTiN film model. We fit the single microwave response measured in transmission using the circle-fit method \cite{probst_efficient_2015}
\begin{equation*}
    S_{21}= 1-\frac{\kappac}{\kappa}\frac{1}{1-2i\Delta/\kappa}
\end{equation*}
and extract the fitting parameters for each value of sheet kinetic inductance (Fig. \ref{fig:nbtinsim}d). 

\label{app:nbtin:parasitic_model}

We find reduced internal quality factor in the coupled system, with $\Qi=4\times10^2-6\times10^3$, compared to independent microwave resonator simulations using identical metal parameters, which yield $\Qi = 10^5$. In the coupled system, the microwave mode acquires additional dissipation associated with coupling to mechanical degrees of freedom. Even away from high-$\Qi$ mechanical resonances, we observe a reduction of the microwave quality factor, which we attribute to a dense spectrum of lossy electromechanical modes \cite{scigliuzzo_phononic_2020}. To quantify this contribution, we use an admittance-based representation of the combined system, in which the internal quality factor of the microwave resonance is given by
\begin{equation*}
    \Qi = \frac{\omega_\mu}{2}\frac{d\Im(Y(\omegamu))}{d\omega}\frac{1}{\Re (Y(\omegamu))}
\end{equation*}
Here, the total admittance is written as a parallel combination of $Y = Y_\mathrm{\mu}+Y_\mathrm{EMC}$, where $Y_\mathrm{\mu}$ is the bare microwave resonator and $Y_\mathrm{EMC}$ captures the electromechanical contribution.

Away from strongly coupled mechanical resonances, $Y_\mathbf{EMC}$ varies slowly with frequency, so that its contribution to $d\Im(Y)/d\omega$ is small compared to that of the bare microwave mode. In this regime, the real part of the EMC admittance  $G_\mathrm{EMC}(\omega)$ acts as a parasitic loss channel whose frequency dependence is smooth but non-negligible. The resulting dissipation can therefore be represented as a frequency-dependent effective shunt resistance $R_p(\omega)=G_\mathrm{EMC}^{-1}(\omega)$. The expression then reduces to 
\begin{equation*}
    Q_p(\omegamu) = \omegamu C_\mathrm{\mu,tot} \frac{R_\mu R_p(\omegamu)}{R_\mu+R_p(\omegamu)}= \omegamu C_\mathrm{\mu,tot} R_p(\omegamu)
\end{equation*}
where $C_\mathrm{\mu,tot} = \frac{1}{2}\frac{d\Im(Y(\omegamu))}{d\omega}$, and $R_\mathrm{\mu}=\frac{\Qmu}{\omegamu C_\mathrm{\mu,tot}}$. This approximation holds in the regime $R_\mathrm{\mu}\gg R_\mathrm{p}(\omegamu)$, where the EMC-induced channel dominates the effective parallel loss. It breaks down near strongly coupled mechanical resonances, where $Y_{EMC}$ varies rapidly and the lumped-element representation is no longer valid. 

The model is shown in Fig. \ref{fig:nbtinsim}d together with the extracted internal quality factors. Away from avoided crossings, it reproduces the frequency-dependent loss in the simulated system, indicating that the background EMC admittance constitutes a finite parasitic dissipation channel even far from the mechanical resonance. The corresponding effective quality factor $Q_p$ is comparable in magnitude to the measured $\Qi$ of the microwave device, showing that this channel constitutes a significant contribution to the overall loss budget in this regime.

\section{Coupled Microwave and Mechanical Mode}
\label{app:s21model}
\label{app:s21model:inputoutput}
The coupling rate is expected to scale with the external coupling rate according to \cite{bosman_approaching_2017, jiang_lithium_2019}
\begin{equation*}
\gem \approx \sqrt{\omegam \omegamu}\frac{\CIDT}{2\sqrt{(\Cmu + \CIDT)(\Cm + \CIDT)}}.
\end{equation*}

At resonance, where $\omegamu= \omegam$ and since the mechanical mode capacitance is much larger than the IDT capacitance, $\Cm\gg \CIDT$, we simplify the expression and substitute $\gammac = \omegam^2\CIDT^2 Z_0/\Cm$, yielding

\begin{equation*}
\gem^2 \approx \frac{\gammac}{4Z_0}\frac{1}{C_\mathrm{\mu,tot}}
\end{equation*}

with $C_\mathrm{\mu,tot}= \Cmu+\CIDT$. 

To obtain an independent estimate of the coupling strength, we characterize reference EMC devices fabricated on the same chip and measured at room temperature. The reference devices share the same design as the EMCs integrated with the microwave resonators. Using the measured external coupling rate, $\gammac/(2\pi)=18\,\mathrm{kHz}$, together with the simulated bare microwave capacitance $\Cmu=10.1\,\text{fF}$, we calculate an expected coupling rate of $\gem/(2\pi)=33\,\mathrm{MHz}$.

To compare the predicted coupling rate with the measured response, we model the microwave transmission of the coupled electromechanical system. We consider a single microwave resonator coupled to a single EMC mode, described by the Hamiltonian

\begin{equation*}
\hat{H}/\hbar = \omegamu \hat{a}^\dagger \hat{a} + \omegam \hat{b}^\dagger\hat{b} + \gem(\hat{a}^\dagger \hat{b}+\hat{a}\hat{b}^\dagger)+c.c.
\end{equation*}

where $\hat{a}$ $(\hat{b})$ is the annihilation operator for the microwave (mechanical) mode \cite{jiang_optically_2023, chiappina_design_2023}.

The Heisenberg–Langevin equations of motion in the frequency domain are

\begin{align*}
0 &= -(i\Delta_\mathrm{\mu} + \frac{\kappa}{2})a+i\gem b-\sqrt{\frac{\kappac}{2}}a_{i,in} \\
0 &= -(i\Delta_\mathrm{m}+\frac{\gamma}{2})b+i\gem a
\end{align*}

with $\kappa$ the total microwave loss, $\kappac$ the external microwave loss rate, $\omegamu$ the microwave resonance frequency and $\gamma$ the total mechanical loss \cite{jiang_lithium_2019}. Applying standard input-output theory yields

\begin{equation*}
a_{i,out} = a_{j,in}-\sqrt{\frac{\kappac}{2}}a
\end{equation*}

Solving these equations for the microwave transmission signal $S_{21} = \frac{a_{2,out}}{a_{1,in}}$ gives

\begin{equation*}
\label{eq:s21}
S_{21} = 1 - \frac{|\kappac|}{\kappa+2i(\omega-\omegamu)+\frac{2\gem^2}{\gamma+2i(\omega-\omegam)}}.
\end{equation*}

To account for tuning we further model the microwave resonance frequency as a current-dependent quantity,
\begin{equation*}
\Delta f / f = -k (I-I_0)^2
\end{equation*}
with $\Delta f$ the frequency shift and $k$ the tuning coefficient. The parameter $I_0$ captures a reproducible offset of the apparent tuning minimum between up- and down-sweeps, consistent with hysteresis-like behavior in the tuning curve. 

We fit our background-corrected data to Eq. (\ref{eq:s21}) using a Markov Chain Monte Carlo optimizer, with $\omegamu$ treated as a current-dependent parameter according to the tuning relation above. The model therefore simultaneously captures both the electromechanical coupling and the magnetic tuning of the mechanical resonance.

\begin{figure}
    \centering
    \includegraphics[width=\linewidth]{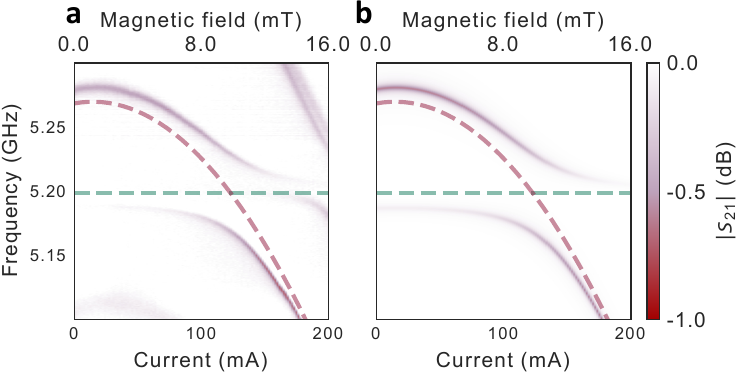}
    \caption{\textbf{Comparison of the measured and simulated microwave transmission magnitude $|S_{21}|$ as a function of bias current.} \textbf{a}, Measure response, same dataset as shown in Fig. \ref{fig:nbtin_emc}. \textbf{b}, Simulated response, calculated from the fitted model parameters.}
    \label{fig:s21fit}
\end{figure}

The fit provides estimates for all model parameters appearing in Eq. (\ref{eq:s21}) and yields an electromechanical coupling rate of $\gem/(2\pi)=30\,\text{MHz}$. This value is in close agreement with the value predicted from the circuit model and independently characterized reference EMCs, confirming the consistency of the fitted electromechanical coupling strength (Fig. \ref{fig:s21fit}).

\begin{table}[h]
\centering
\caption{\textbf{NbTiN-EMC device parameters}}
\begin{tabular}{ll}
\toprule
\textbf{Parameter} & \textbf{Value} \\
\midrule
$f_\mathrm{\mu}$      & $5.27\,\text{GHz}$ \\
$f_\mathrm{m}$      & $5.20\,\text{GHz}$ \\
$Q_\mathrm{\mu,i}$ & $1.2 \times 10^3$ \\
$Q_\mathrm{m}$ & $1.2\times 10^3$\\
$Q_\mathrm{\mu,c}$ &  $1.4\times 10^4$ \\ 
$\Zmu$ & $2.5\, \mathrm{k\Omega}$ \\
$ Z_\mathrm{\mu,tot}$ & $2.2\,\mathrm{k\Omega}$ \\
$\Cmu$ & $10.1\,\text{fF}$ \\
$C_\mathrm{\mu,tot}$ & $13.4\,\text{fF}$ \\

$L_\mathrm{k}$ & $64.0\,\text{nH}$ \\
$L_\mathrm{g}$ & $0.2\,\text{nH}$ \\
$\sigma_\mathrm{g}$ & $1.9\,\text{sq}$ \\
$\CIDT$ & $3.3\,\text{fF}$ \\
$\gem/(2\pi)$ & $30\,\text{MHz}$ \\
$\mathcal{C}_\mathrm{em}$ & 180\\
\bottomrule
\end{tabular}
\end{table}

\section{EMC Loss Sources}
\label{app:loss}
To estimate the expected loss rates of our device across temperature and intracavity occupation, we compare against prior work using two complementary approaches: (i) a participation-ratio-weighted decomposition of material losses and (ii) direct benchmarking against intrinsic quality factors reported for geometrically similar resonators. In particular, we consider lithium niobate surface acoustic wave (SAW) devices and suspended phononic crystal (PnC) defect resonators as relevant reference systems \cite{wollack_loss_2021,arrangoiz-arriola_resolving_2019}, as both share key features with our architecture, including strong piezoelectric coupling and partial energy localization in electrode and surface regions.

Although these device classes differ in geometry and confinement mechanism, both exhibit internal quality factors in the range relevant for the present devices and are subject to comparable loss channels arising from surface participation, radiation into the substrate, and electrode-induced dissipation. In SAW devices, acoustic energy is primarily confined near the surface and can radiate into the substrate through imperfect phononic band gap confinement and surface scattering. In suspended phononic crystal defect cavities, while radiation loss through finite band gap attenuation and imperfect confinement contributes to dissipation, experimental evidence in lithium niobate systems indicates that surface-related TLS loss and strain participation in metallic electrodes can provide comparable or dominant contributions, particularly in nanoscale geometries with large surface-to-volume ratio \cite{wollack_loss_2021}. Both platforms provide useful benchmarks for order-of-magnitude estimates of internal quality factors in lithium niobate-based electromechanical systems.

\label{app:loss:rt}
At room temperature, dissipation is dominated by intrinsic material damping in both the piezoelectric substrate and the metallic electrodes. In particular, the aluminum electrodes exhibit comparatively large mechanical loss due to grain-boundary and defect-mediated dissipation, which is known to limit the performance of electromechanical resonators across a wide range of geometries \cite{ryzy_measurement_2018}. Reported effective mechanical quality factors for aluminum films in resonant structures are typically in the range $\Qi\sim200-300$ \cite{schaffer_measurement_2023, grunsteidl_measurement_2020}, while lithium niobate-based devices with reduced electrode participation can reach $\Qi \sim 950$ \cite{jiang_efficient_2020} at \SI{1.8}{\GHz}. For comparison, SAW resonators on lithium niobate have demonstrated room-temperature quality factors up to $\Qi \sim 2.5\times 10^3$ at \SI{5}{\GHz} \cite{shao_phononic_2019}.

In our electromechanical cavities (EMCs), the fraction of elastic energy stored in the aluminum electrodes is estimated from finite-element simulations to lie in the range $12-20\% $, depending on electrode thickness and mode profile. Using a representative participation ratio of $15\% $ and assigning effective loss-limited quality factors of 300 for aluminum and 1500 for lithium niobate, the resulting participation-weighted estimate yields a room-temperature limit of $\Qi \sim 940$. This estimate is consistent with previously reported aluminum–lithium niobate hybrid devices, which typically exhibit room-temperature quality factors in the range $\Qi \sim 300-400$ \cite{sarabalis_s-band_2020, mayor_gigahertz_2021}. Our devices fall within this range, with typical values of $\Qi\sim 200-500$, and a maximum observed value of $\Qi=810$, consistent with the expected loss-limited regime.

\begin{figure}
    \centering
    \includegraphics[width=\linewidth]{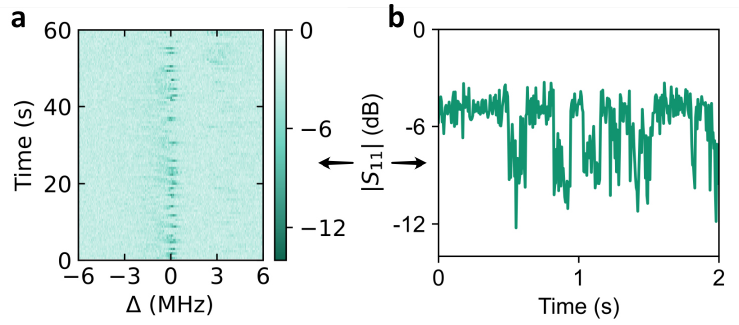}
    \caption{\textbf{Telegraphing behavior of main mechanical mode in LNOS EMCs at low phonon occupation $n_\mathrm{ph}<1$.} \textbf{a} Microwave reflection measurement swept over the mechanical resonance found at high power ($\Delta = \omega-\omegam$). Low drive powers are used to reach $n_\mathrm{ph}<1$ . \textbf{b}, Time trace of the microwave reflection signal at zero detuning ($\Delta=0$), showing telegraphic switching.}
    \label{fig:telegraphing}
\end{figure}

\label{app:loss:cryo}
At cryogenic temperatures, reliable quantitative data on material loss in lithium niobate hybrid systems remains limited, and we therefore benchmark against experimentally reported device performance. In this regime, dissipation depends strongly on both geometry and excitation level, reflecting the interplay between radiation loss mechanisms and two-level system (TLS) defects at surfaces and interfaces.

Representative lithium niobate SAW and phononic crystal resonators exhibit internal quality factors in the range $\Qi\sim 10^4-10^5$, depending on device geometry and drive conditions. Phononic crystal defect cavities with aluminum electrodes have demonstrated high-power quality factors approaching $\Qi\sim 10^5$ \cite{wollack_loss_2021}, while SAW devices typically operate in the range $\Qi \sim 10^4-10^5$ at comparable frequencies \cite{shao_phononic_2019, luschmann_surface_2023}. At lower excitation levels, a reduction in quality factor is commonly observed, with reported values down to $\Qi\sim 10^4$ in both SAW and suspended geometries, consistent with increased TLS participation \cite{gruenke_surface_2024}.

For reference, Table \ref{tab:Qi} summarizes representative cryogenic quality factors across lithium niobate-based electromechanical systems, including SAW devices, suspended phononic crystal resonators, and related hybrid geometries.

\begin{table*}[ht]
\centering
\begin{threeparttable}
\caption{\textbf{Representative internal quality factors $\Qi$ and coupling parameters of lithium niobate-based electromechanical resonators at cryogenic temperatures.} “High power” refers to regimes in which TLS saturation effects are reduced due to large intracavity phonon occupation, while “low power” corresponds to low-excitation operation ($n_\mathrm{ph}<1$) where TLS-induced loss is typically enhanced. Here, $\gamma_c$ denotes the external coupling rate and $g_\mathrm{em}$ the electromechanical coupling rate. Abbreviations: non-suspended phononic racetrack (NS-PCR), suspended phononic crystal (S-PC), and non-suspended phononic crystal (NS-PC). }
\label{tab:Qi}
\begin{tabular}{llllllll}
\toprule
\textbf{Source} & Device & $\omegam/(2\pi)$ (GHz) & $\gammac/(2\pi)$  & $\gem/(2\pi)$ & High power $\Qi$ & Low power $\Qi$ & Year\\
\midrule
\cite{gruenke_surface_2024}* & SAW & 0.7 & \SI{70}{\kHz}& - & $1.5 \times 10^4$ & $10^4$ & 2024 \\
\cite{luschmann_surface_2023} & SAW & 4.8 & \SI{134}{\kHz} & - & $8.5 \times 10^4$ & $6.4\times10^4$ & 2023\\
\cite{shao_phononic_2019} & SAW & $1.0$ & - & - & $6.1 \times10^4$& - & 2019\\
\cite{wang_circuit_2025} & NS-PCR & $3.9$ & -& \SI{0.36}{\MHz}& $2.1\times 10^4$ & $1.2 \times 10^4$ & 2025\\
\cite{mayor_gigahertz_2021} & NS-PCR & $3.4$ & \SI{50}{\MHz} & - &$4.7\times 10^4$ & - & 2021\\
\cite{wollack_quantum_2022} & S-PC & 2.0 & - & \SI{10.5}{\MHz} & - & $1.6\times 10^4$ & 2022\\
\cite{wollack_loss_2021}** & S-PC & $2.0$ & \SI{85}{\kHz} & - & $10^5$ & $7.0\times 10^3 $ & 2021 \\
\cite{jiang_lithium_2019} & S-PC & 2.0 & \SI{8.8}{\mHz} & - & $1.7\times10^4$ & - & 2019\\
\cite{arrangoiz-arriola_resolving_2019} & S-PC & 2.4 & - & \SI{15.7}{\MHz} & - & $6.5\times 10^3$ & 2019\\
LNOS EMC (This work) & NS-PC & 4.7 & \SI{72.6}{\kHz} & \SI{30}{\MHz}& $10^4$ & $7.0\times10^3$ & 2026\\
LiSa EMC (This work) & NS-PC & 5.2 & \SI{198}{\kHz} & - & $10^4$ & $8.5\times10^3$ & 2026\\
\bottomrule
\end{tabular}
\vspace{1mm}
{\footnotesize
$^{*}$ Measured on two separate devices.\\
$^{**}$ For aluminum on-defect devices, with estimated low power $\Qi$.
}
\end{threeparttable}
\end{table*}

Our electromechanical cavities operate around \SI{5}{\GHz} and exhibit internal quality factors of $\Qi\sim 10^4$ at high excitation and $\Qi\sim 7.0-8.5\times 10^3$ at low excitation. These values are below those reported for the highest-performing lithium niobate SAW and phononic crystal resonators, but fall within the broader range observed across lithium niobate electromechanical devices.

The observed performance is consistent with a regime in which dissipation arises from a combination of radiation loss and material damping, with additional contributions from TLS-mediated processes at low excitation. In particular, the reduction in quality factor at low phonon occupation is consistent with previous observations of TLS-induced spectral diffusion and telegraphic switching in suspended phononic crystal resonators at millikelvin temperatures \cite{maksymowych_spectral_2025}. This is further supported by the observed telegraphic switching behavior in the microwave response of our devices at $n_\mathrm{ph}<1$ (Fig. \ref{fig:telegraphing}), which limits stable fitting of the resonance lineshape due to time-dependent frequency fluctuations. Across regimes, dissipation evolves from metal- and phonon–phonon-dominated damping at room temperature, to a radiation- and metal-loss-limited regime at cryogenic high excitation, and finally to a mixed regime at low excitation where TLS-induced fluctuations coexist with residual radiation loss.

These observations motivate further optimization of electrode geometry and surface processing to reduce both radiation and interface participation losses, with the goal of approaching the upper range of reported lithium niobate electromechanical resonator performance ($\Qi \sim 10^5$ at high excitation).

\section{EMC Temperature Dependence}
\label{app:Tsweep}
To determine the temperature-dependent behavior of our devices we measure the microwave reflection spectra $S_{11}$ of a LiSa EMC over the course of a cool down. The data is acquired and fitted following the procedure described in Sec. \ref{sec:cryos11}, using high microwave drive power. The presented data correspond to an EMC geometry outside the sweep shown in Fig.~\ref{fig:RTs11}e. We present data between 100 K to 4 K, as the mixing chamber temperature is not monitored above 100 K. At higher temperatures, increased material losses further reduce the signal strength, while a larger measurement background than in the room-temperature characterization makes reliable fitting increasingly difficult.

\begin{figure}
    \centering
    \includegraphics[width=\linewidth]{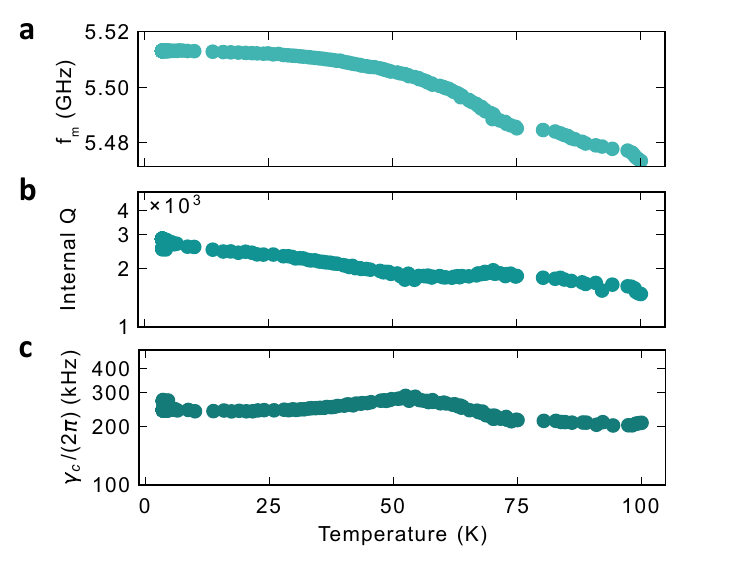}
    \caption{\textbf{Temperature dependent performance of LiSa EMC.} \textbf{a}, Mechanical resonance frequency of a LiSa EMC as a function of temperature, showing a gradual increase upon cooling. \textbf{b}, Internal quality factor extracted from fits to the microwave reflection spectra. \textbf{c}, External coupling rate extracted from the same measurements. }
    \label{fig:Tsweep}
\end{figure}

The mechanical resonance frequency gradually increases as the device is cooled, saturating below approximately \SI{25}{\kelvin}. The resonance frequency increases from \SI{5.45}{\GHz} at room temperature by approximately $1.1\%$, consistent with frequency shifts arising from temperature-dependent changes in the elastic constants of lithium niobate \cite{shao_phononic_2019}.

Over the measured range, the internal quality factor increases gradually from approximately 1000 to 3000 for the selected device. The measured device exhibited a room-temperature internal quality factor of $(\Qi = 690)$, among the highest observed LiSa EMCs at room temperature. However, several other measured devices achieved higher $\Qi$ at cryogenic temperatures, suggesting that room-temperature $\Qi$ is not a reliable predictor of cryogenic performance.

Over the measured temperature range, the external coupling rate $\gammac$ exhibits a weak temperature dependence, from $\gammac\sim 200 \,\text{kHz}$ at 100 K to $\gammac\sim 250 \,\text{kHz}$ at 4 K (Fig. \ref{fig:Tsweep}c), compared to \SI{150}{\kHz} measured in the room temperature setup. In the absence of a sharp transition (e.g. due to onset of superconductivity), we attribute the gradual increase in coupling rate to changes in the microwave measurement environment.

\section{Cryogenic Measurement Setup}
\label{app:cryo_setup}
\begin{figure}[h!]
    \centering
    \includegraphics[width=\linewidth]{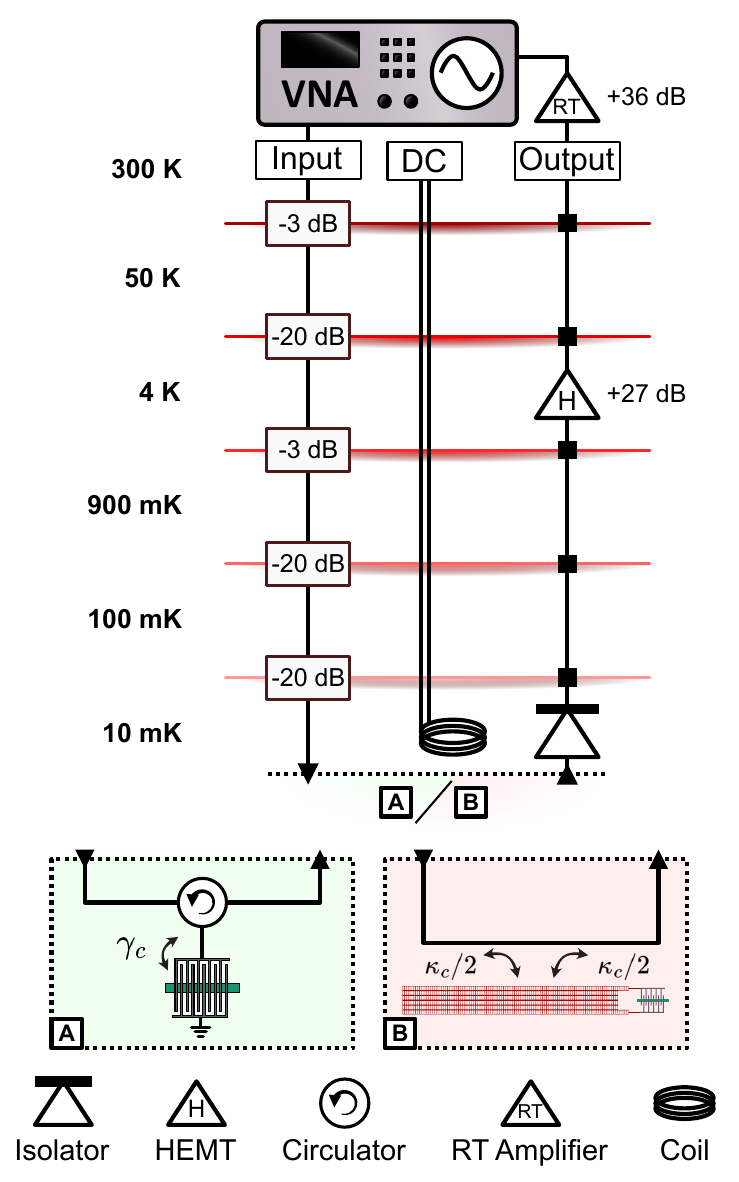}
    \caption{\textbf{Experimental setup for measurement at cryogenic temperatures.}The sample is mounted on a microwave PCB inside a dilution refrigerator, with an external magnetic coil and magnetic shielding. For reflection measurements of individual EMC devices, a circulator is used (A).For high-impedance microwave resonators, transmission measurements are performed directly through the PCB (B).}
    \label{fig:cryo_setup}
\end{figure}

We perform all measurements using a Rohde \& Schwarz ZNB20 vector network analyzer (VNA). The sample is wire-bonded to a PCB providing up to 8 electrical connections and hosting two chips. For bare EMC devices, a microwave switch is used to sequentially measure up to six devices. For the coupled electromechanical system, we use a chip layout with transmission feedlines coupled to multiple microwave resonators at different frequencies (Fig.~\ref{fig:cryo_setup}). 

The input line is attenuated by a total of 66 dB using distributed cryogenic attenuators. We measure the attenuation through the input line and an output line with identical cabling, without attenuators on the output line. After accounting for the known attenuators, the remaining one-way cable attenuation is found to vary approximately linearly with frequency, increasing from 6.5 dB at \SI{4}{\GHz} to 8.5 dB at \SI{6}{\GHz}. Including insertion losses from the microwave switch (Radiall R591763600) and the circulator (LNF-CiC$4\_12$A), the total attenuation at the device plane is approximately 74 dB at \SI{5}{\GHz}. The uncertainty in the absolute power calibration is estimated to be approximately $\pm1-2\,\text{dB}$, primarily arising from differences between the input and output cable assemblies and connector losses associated with the cryogenic attenuators.

The reflected signal is amplified at the 4 K stage using a high-electron mobility transistor (HEMT) amplifier (LNF-LNC$0.3\_14$B), followed by additional room-temperature amplification (LNF-LNR$1\_15$B$\_$SV), yielding a total nominal gain of approximately 63 dB.

The HEMT is well described by an equivalent input-referred noise temperature of approximately $3-4$ K and dominates the output noise performance. Back-propagating noise from higher-temperature stages is suppressed by the cryogenic microwave chain. In particular, noise emitted by the HEMT is attenuated at the sample by $\sim20$ dB isolation of the mixing-chamber circulator, which provides the dominant reverse isolation between the 4 K stage and the device. This results in an estimated effective noise temperature at the sample on the order of $T\sim40 \,\mathrm{mK}$, corresponding to a thermal occupation of $n_\mathrm{th} \sim 2\times 10^{-3}$ at \SI{5}{\GHz}.

A DC current of 0–200 mA (Yokogawa GS200) is supplied to an external coil to generate the magnetic field. Rapid changes in current lead to measurable heating, which we attribute to eddy-current dissipation in nearby metallic structures. To maintain thermal stability, the current is ramped slowly at a rate of \SI{0.4}{\milli\ampere\per\second}.

\section{EMC Length and Electrode Design Considerations}
\label{app:len}

\label{app:len:overview}
The EMC geometry is defined by the number of mirror, transition, and defect cells, as well as the extent of IDT electrode coverage. Defect and transition cells determine the effective cavity length, while mirror cells primarily control confinement. These parameters influence the mechanical quality factor, electromechanical coupling strength, mode density, and fabrication yield, leading to several competing design trade-offs. Electrode coverage provides an additional degree of freedom that primarily affects the electromechanical coupling and associated scattering losses.

\begin{figure*}[ht!]
    \centering
    \includegraphics[width=\linewidth]{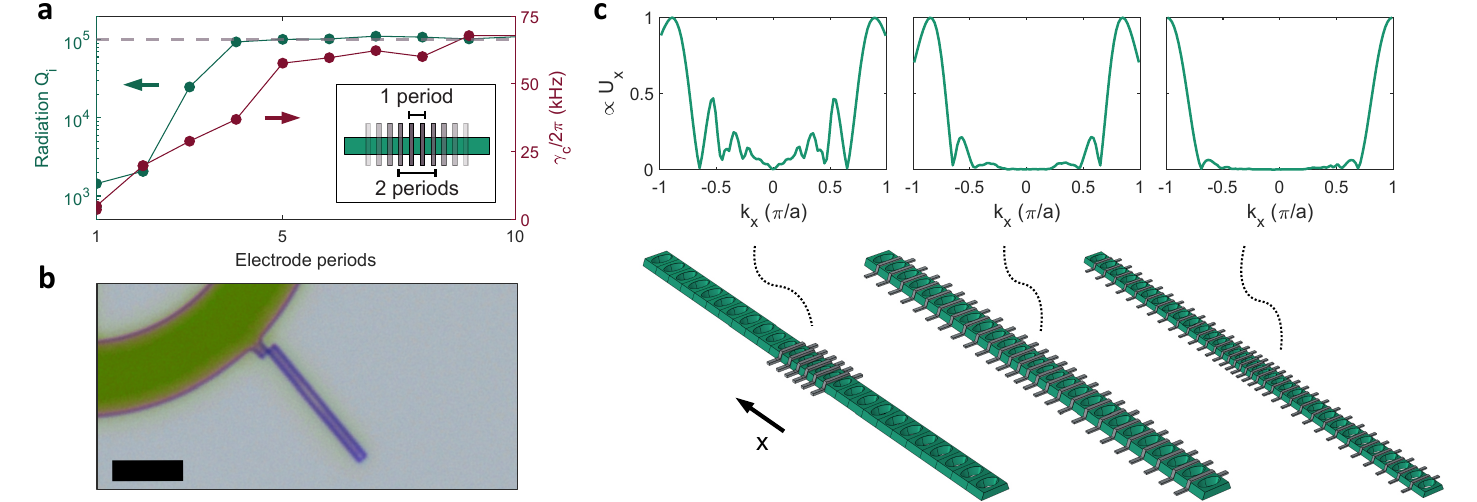}
    \caption{ \textbf{Effects of length scaling of the EMC} \textbf{a}, Internal quality factor $\Qi$ and external coupling rate $\gammac$ of an optimized EMC versus number of electrode periods. \textbf{b}, Micro-graph of an EMC broken during the micro-transfer printing fabrication step (scale bar \SI{5}{\um}). \textbf{c}, Fourier spectra of the displacement field of three EMC geometries demonstrating the effect of cavity length and electrode coverage. Shown for normalized displacement amplitude at k-vectors expressed in units of $\pi/a$ with $a$ the defect cell period.} 
    \label{fig:fourier}
\end{figure*}
\label{app:len:cell_roles}
The electromechanical coupling depends on the spatial overlap between the mechanical displacement field and the electrode coverage. A larger electrode-covered region increases the total IDT capacitance approximately linearly with the number of electrode periods, $C_0 \approx NC_\mathrm{0,n}$, where $C_\mathrm{0,n}$ denotes the average capacitance per electrode period. Assuming a uniform mechanical displacement beneath all electrodes, the electromechanical coupling then follows the ideal scaling $\gem\propto \sqrt{N}$.

Adding defect cells approximately preserves the displacement amplitude over the electrode region, enabling coupling that approaches this ideal scaling. In contrast, additional transition cells extend the spatial taper of the mode, reducing the average displacement amplitude per added electrode period and resulting in sub-$\sqrt{N}$ scaling. Thus, the effective coupling depends not only on electrode number but also on how the cavity length is distributed between defect and transition regions.

\label{app:len:electrodes}
To isolate the effect of electrode coverage, we optimize an EMC with the same unit cell configuration as the main device in Sec. \ref{sec:design} while varying the number of IDT electrode periods (Fig. \ref{fig:fourier}a).

Electrode termination introduces a perturbation to the periodic structure, leading to scattering and a reduction in piezoelectric interaction when significant mechanical amplitude remains at the truncation point. However, when the mechanical mode is sufficiently attenuated, additional electrode periods no longer contribute appreciably to either confinement or coupling.

We find that beyond approximately five electrode periods, both the radiation-limited quality factor $\Qi$ and external coupling rate $\gammac$ saturate, indicating that the mode is effectively confined within the electrode-covered region. Additional electrode periods beyond this point primarily increase parasitic capacitance and improve robustness to fabrication disorder, but provide diminishing returns in terms of device performance.

\label{app:len:fourier}
In addition to affecting electromechanical coupling, the distribution of defect and transition cells also determines the confinement of the mechanical mode and therefore the radiation-limited quality factor. To connect the observed design trends to radiation loss mechanism, we analyze the Fourier spectra of the mechanical displacement field for different EMC geometries (Fig. \ref{fig:fourier}c).

Long cavities exhibit a narrower k-space distribution with reduced weight at low wave vectors, increasing their separation from the substrate acoustic continuum. A similar effect is observed for extended electrode coverage, where electrode termination acts as an additional perturbation to the mode profile.

To suppress radiation loss, it is therefore desirable to minimize overlap with low-k components that couple to propagating surface acoustic waves (SAWs). The SAW continuum begins at approximately $0.53\,\pi/a$ in silicon ([100]), $0.46 \,\pi/a$ in C-plane sapphire, and $0.24\,\pi/a$ in diamond. The choice of substrate therefore shifts the position of the radiation continuum, in addition to modifying the mode profile.

The suppression of low-k components is strongest for long cavities with electrode coverage extending across the full defect and transition region. In simulation, a cavity with 11 transition cells and 1 defect cell reaches a radiation-limited quality factor of $\Qi=5\times10^5$. However, as discussed above, fabricated devices are expected to be limited below this value.

\label{app:len:mode_density}
Increasing the cavity length increases the number of localized mechanical modes supported within the frequency range of interest, resulting in a higher mode density. As a consequence, the spectral separation between modes is reduced, making individual mode isolation more challenging and increasing the likelihood of unwanted coupling to nearby mechanical modes, including Purcell-like decay processes. Further discussion of release-free mode density can be found in the supplementary of \cite{burger_design_2025}.

\label{app:len:fab}
The cavity length also influences fabrication yield during the micro-transfer printing process, where the EMC is attached at one end and freely suspended at the other. The weakest regions of the structure are typically located in the mirror cells, which have a smaller cross-section due to their larger holes compared to the defect cells.

This is reflected in fabricated devices, where fractures predominantly occur in the mirror region close to the coupon (Fig. \ref{fig:fourier}b). Increasing the suspended length increases the mechanical compliance of the structure during handling, making fracture at these narrow regions more likely. We therefore attribute these fracture events to mechanical stresses experienced during micro-transfer printing, with the probability of fracture increasing for longer suspended structures.

\label{app:len:summary}
Increasing cavity length improves confinement and electromechanical coupling, but increases mode density and reduces fabrication yield. Electrode coverage provides an additional optimization parameter, where sufficient coverage is required to avoid electrode-termination scattering, but excessive coverage provides diminishing returns. The final EMC design therefore represents a compromise between coupling strength, mechanical quality factor, mode isolation, and fabrication robustness.

\end{document}